\documentclass[aps,prb,twocolumn,superscriptaddress,floatfix,longbibliography]{revtex4-2}

\usepackage{graphicx,graphics}
\usepackage{dcolumn}
\usepackage{amsmath}
\usepackage{amsfonts}
\usepackage{amssymb}
\usepackage{latexsym,verbatim}
\usepackage{physics}
\usepackage{bm}
\usepackage{color}
\usepackage{xcolor}
\usepackage[normalem]{ulem}
\usepackage[percent]{overpic}
\usepackage{bbold}
\usepackage{lipsum}
\usepackage{ragged2e}
\usepackage[breaklinks=true,colorlinks,citecolor=blue,linkcolor=blue,urlcolor=blue]{hyperref}

\usepackage{graphicx}

\DeclareMathAlphabet\mathbfcal{OMS}{cmsy}{b}{n}

\newcommand{\kF}{k_{\rm F}}
\newcommand{\vD}{v_{\mathrm{D}}}

\newcommand{\epsz}{\epsilon_{\mathrm{Z}}}

\usepackage{soul}
\usepackage{orcidlink}

\begin{document}

\title{Local analysis of a single impurity on a graphene Josephson Junction}
\author{Ignazio~Vacante}
\affiliation{Dipartimento di Fisica e Astronomia "Ettore Majorana",  Universit\'a di Catania, Via S. Sofia, 64, I-95123 Catania,~Italy.}
\affiliation{INFN, Sez. Catania, I-95123 Catania,~Italy.}
\author{Francesco M.D.~Pellegrino~\orcidlink{0000-0001-5425-1292}}
\email{francesco.pellegrino@dfa.unict.it}
\affiliation{Dipartimento di Fisica e Astronomia "Ettore Majorana",  Universit\'a di Catania, Via S. Sofia, 64, I-95123 Catania,~Italy.}
\affiliation{INFN, Sez. Catania, I-95123 Catania,~Italy.}
\affiliation{CNR-IMM, Catania (University) Unit, via Santa Sofia 64, I-95123 Catania,~Italy.}
\author{G.G.N.~Angilella~\orcidlink{0000-0003-2687-9503}}
\affiliation{Dipartimento di Fisica e Astronomia "Ettore Majorana",  Universit\'a di Catania, Via S. Sofia, 64, I-95123 Catania,~Italy.}
\affiliation{Scuola Superiore di Catania, Universit\'a di Catania, 9, Via Valdisavoia, I-95123 Catania, Italy.}
\affiliation{INFN, Sez. Catania, I-95123 Catania,~Italy.}
\affiliation{Centro Siciliano di Fisica Nucleare e Struttura della Materia, Catania, Italy.}
\author{Giuseppe A.~Falci~\orcidlink{0000-0001-5842-2677}}
\affiliation{Dipartimento di Fisica e Astronomia "Ettore Majorana",  Universit\'a di Catania, Via S. Sofia, 64, I-95123 Catania,~Italy.}
\affiliation{INFN, Sez. Catania, I-95123 Catania,~Italy.}
\affiliation{CNR-IMM, Catania (University) Unit, via Santa Sofia 64, I-95123 Catania,~Italy.}
\author{Elisabetta~Paladino~\orcidlink{0000-0002-9929-3768}}
\affiliation{Dipartimento di Fisica e Astronomia "Ettore Majorana",  Universit\'a di Catania, Via S. Sofia, 64, I-95123 Catania,~Italy.}
\affiliation{INFN, Sez. Catania, I-95123 Catania,~Italy.}
\affiliation{CNR-IMM, Catania (University) Unit, via Santa Sofia 64, I-95123 Catania,~Italy.}

\begin{abstract}
In this work, we investigate the local effects of a single impurity on the electron system of a short ballistic graphene Josephson Junction. Within the Dirac-Bogoliubov-de Gennes approach, we systematically analyze the local electron density of states. Its behavior at subgap energies allows one to distinguish elastic and inelastic scattering processes and to identify the magnetic nature of the impurity. Moreover, the spatial dependence of the local density of states is a sensitive probe of microscopic processes yielding subgap impurity-induced bound states. The Fourier analysis allows one to recognize the wavevectors related to the momenta of the high-transmissive channels in ballistic graphene.
\end{abstract}

\maketitle

\section{Introduction}

Hybrid systems that combine different physical platforms with complementary functionalities are nowadays explored for applications in quantum technologies~\cite{Laucht_2021_Nanotech,deleon_science_2021}.
For instance, gate-tunable hybrid superconducting qubits known as gatemons have been implemented using semiconducting nanowires~\cite{larsen_prl_2015, delange_prl_2015}, InAs-based Josephson Junctions (JJs)~\cite{nichele_prl_2020,Kringhoj_PRB_2018,Kringhoj_PRL_2020}, two-dimensional materials~\cite{Lee_NanoLett_2019}, van der Waals heterostructures~\cite{Abhinandan_NanoLett_2021}, and graphene~\cite{Casparis_NatNanotech_2018, wang_natnanotech_2019}. They offer advantages such as reduced dissipative losses and crosstalk, as well as compatibility with high magnetic fields~\cite{schmidt_natcomm_2018,kroll_natcomm_2018}, paving the way for the implementation of fault-tolerant topological qubits based on Majorana zero modes~\cite{shabani_prb_2016,Nayak_RevModPhys_2008}. High-quality graphene-superconductor heterostructures with clean interfaces obtained by encapsulating graphene in hexagonal boron nitride (hBN)~\cite{dean_natnano_2010,mayorov_nanolett_2011, wang_science_2013} sustain ballistic transport of Cooper pairs over micrometer-scale lengths, gate-tunable supercurrents persisting at large parallel magnetic fields~ \cite{calado_natnanotech_2015,benshalom_natphys_2016,borzenets_prl_2016}, and show various features of two-dimensional Andreev physics~\cite{allen_natphys_2016,amet_science_2016,bretheau_natphys_2017}. Moreover, thanks to the low specific heat of hBN-embedded graphene highly sensitive microwave bolometers based on graphene Josephson Junction (GJJ)
have been demonstrated for circuit quantum electrodynamics applications~\cite{lee-efetov_nat_2020,kokkoniemi_nat_2020}. Single-photon detection at near-infrared frequencies has also been achieved by coupling photons to localized surface plasmons of a GJJ, which can be integrated into JJ-based quantum architectures as high-speed, low-power optical interconnects~\cite{walsh-efetov_science_2021}. GJJs are also an excellent platform for generating exotic quantum states, such as long-lived Floquet-Andreev states obtained by applying continuous microwave light without significant heating~\cite{park_nature_2022}.
Recently, a superconducting circuit has been manufactured incorporating three Josephson junctions based on graphene governed by a dominant $\sin(2 \phi)$  current-phase relation (CPR)~\cite{messelot_prl_2024}, which is promising for the development of superconducting quantum bits protected from decoherence.

The Josephson effect in heterostructures is due to the proximity effect in combination with constructive interference between Andreev processes at the two normal-superconducting (N-S) interfaces. This results in the formation of coherent electron-hole superpositions known as Andreev bound states (ABSs). In short junctions, the CPR resulting from the phase dependence of the ABSs spectrum differs from the sinusoidal CPR of standard tunnel JJs~\cite{golubov_rmp_2004, Bretheau_PRX_2013}.
Recently, there has been interest in controlling spatial properties of ballistic GJJ. An experimental method has been developed to measure and control the current density in real-space for two parallel ballistic GJJs, based on the analysis of the Fourier and Hilbert transformations of the modulation of the critical current induced by a magnetic field~\cite{schmidt_prapplied_2023}.
Tunneling spectroscopy measurements in GJJs revealed the possible presence of weakly coupled microscopic quantum dots to proximitized graphene, which act as energy filters in the tunneling process~\cite{wang_natnanotech_2019,wang_prb_2018}.
A theoretical analysis has suggested that a homogeneous dilute distribution of impurities in a GJJ can alter the critical current and the skewness of the CPR~\cite{pellegrino_commphys_2022}.
Moreover, defects in the nearby substrate can act as carrier traps and induce fluctuations in the carrier density of the graphene channel~\cite{pellegrino_jstat_2019,kumar_apl_2021} leading to critical current fluctuations with a $1/f$ spectrum~\cite{pellegrino_commphys_2020,pellegrino_epjst_2021,paladino_rmp_2014} and relaxation~\cite{spagnolo_relaxation_2012}.
%
In this work, we investigate the effect of a short-range impurity on the low-energy properties of a short planar GJJ, motivated by findings related to point systems weakly interacting with proximitized graphene and by
the recent interest in controlling the spatial properties of ballistic GJJs~\cite{schmidt_prapplied_2023}. We use an analytical approach based on the Dirac-Bogoliubov-de Gennes model~\cite{titov_prb_2006}. Two paradigmatic descriptions of a single impurity are considered: the Anderson model~\cite{anderson_prev_1961} and the Lifshitz model~\cite{lifshitz_book}. The latter describes an impurity interacting with the density of conduction electrons through a scalar potential~\cite{balatsky_rmp_2006,bespalov_prb_2018} and it has been used to model a vacancy in a graphene crystal~\cite{pellegrino_prb_2009}. The Anderson model has been used to study the effect of adatoms on the graphene electron system~\cite{farjam_prb_2011,yuan_prb_2010,wehling_prl_2010,skrypnyk_jpcondmat_2013,barth_prb_2022}. It allows electron transfer from the host to some energy level belonging to the adsorbed atom~\cite{skrypnyk_jpcondmat_2013}.
In both cases, we take into account an extra magnetic term describing the exchange interaction between the local spin on the impurity site and the electrons flowing through the GJJ.
We focus on a single impurity located in the middle of the normal region and analyze how it affects the local density of states (LDOS) within the superconductive gap $ |E|<\Delta$. This quantity is experimentally accessible by scanning tunneling spectroscopy~\cite{balatsky_rmp_2006} the LDOS
unveiling sensitivity of the ABSs to the type of impurity-scattering processes.
Close to the impurity, the energy dependence of the LDOS is sensitive to the magnetic character of the impurity itself, and it is different for the Anderson and the Lifshitz models.
We analyze the conditions for the appearance of an impurity-induced subgap bound state. Finally, we analyze the spatial dependence in the normal phase region along the transverse direction at the energy corresponding to the impurity-induced bound state.

This paper is structured as follows. Section~\ref{sect:model} provides technical details on the Dirac-Bogoliubov-De Gennes model describing ABSs. Sections~\ref{sect:anderson} and~\ref{sect:lifshitz} present the Anderson and Lifshitz models, respectively, and the analysis of the energy dependence of the LDOS close to an impurity in the middle of the normal phase region. Section~\ref{sect:realspace} addresses the spatial dependence of the LDOS along the normal phase region, and the analysis in Fourier space. Finally, the conclusions are drawn in Section~\ref{sect:conclusions}.

\section{Model}
\label{sect:model}

\begin{figure}[t]
\centering
\begin{overpic}[width=\columnwidth]{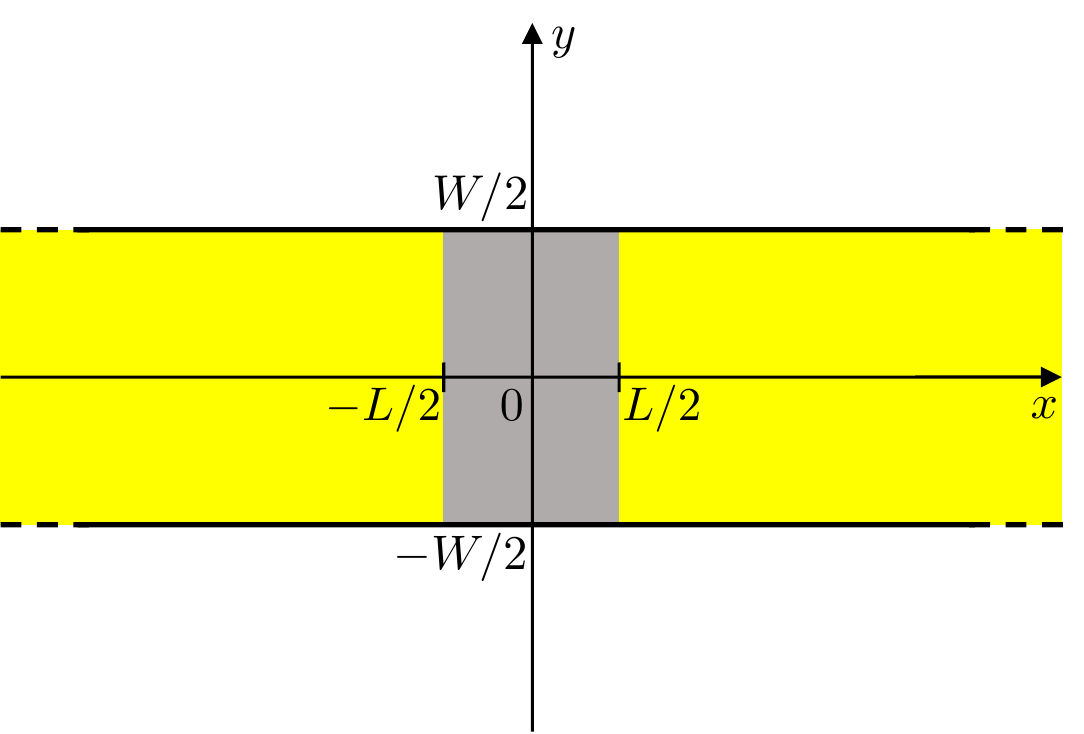}\end{overpic}\vspace{0em}
\caption{\justifying Schematics of the planar GJJ device.
The graphene monolayer of length $L$ (gray) is contacted to the sides by two superconducting electrodes (yellow sides).
The width $W$ of the device is assumed to be such that $W \gg L$. \label{fig:GJJ}}
\end{figure}

We analyze a GJJ in the ballistic regime, as depicted in Fig.~\ref{fig:GJJ}, which is composed of two identical metallic superconductive electrodes on the sides and a graphene monolayer in the middle.
The normal graphene stripe has length $L$ in the $\hat{\bm x}$-direction, the entire device has a finite width $W$ and is considered infinite along the $\hat{\bm x}$-direction, and we assume that $W\gg L$.
The electron system is described by the following Dirac-Bogoliubov-de Gennes (D-BdG) Hamiltonian~\cite{titov_prb_2006}
\begin{equation}\label{eq:DBdG_Hamiltonian}
\begin{aligned}
\hat{\mathcal{H}}_{\mathrm{D}-\mathrm{BdG}}=&\sum_{\zeta= \pm} \int d^{2} \boldsymbol{r} \hat{\Psi}_{\zeta}^{\dagger}(\boldsymbol{r}) H_{\mathrm{D}-\mathrm{BdG}} \hat{\Psi}_{\zeta}(\boldsymbol{r}),\\
H_{\mathrm{D}-\mathrm{BdG}}=& \tau_{z}\left[U(\boldsymbol{r}) \mathbb{1}_{\sigma}+\frac{\hbar v_{\mathrm{D}}}{i}\left(\partial_{x} \sigma_{x}+\partial_{y} \sigma_{y}\right)\right]\\
&+\tau_{x} \mathbb{1}_{\sigma} \operatorname{Re} \Delta(\boldsymbol{r})-\tau_{y} \mathbb{1}_{\sigma} \operatorname{Im} \Delta(\boldsymbol{r})~,
\end{aligned}
\end{equation}
where $\zeta=\pm$ denotes valley indices.
Here, the four components spinors $\hat{\Psi}_{\pm}(\boldsymbol{r})$ are defined as
\begin{equation}\label{eq:spinors}
{\small
\begin{aligned}
    &\hat{\Psi}_{+}(\boldsymbol{r})=\left[\hat{\psi}_{A, \boldsymbol{K}, \uparrow}^{\dagger}(\boldsymbol{r}), \hat{\psi}_{B, \boldsymbol{K}, \uparrow}^{\dagger}(\boldsymbol{r}), \hat{\psi}_{A, \boldsymbol{K}^{\prime}, \downarrow}(\boldsymbol{r}), \hat{\psi}_{B, \boldsymbol{K}^{\prime}, \downarrow}(\boldsymbol{r})\right]^{\dagger}~,\\
    &\hat{\Psi}_{-}(\boldsymbol{r})= \left[-\hat{\psi}_{B, \boldsymbol{K}^{\prime}, \uparrow}^{\dagger}(\boldsymbol{r}), \hat{\psi}_{A, \boldsymbol{K}^{\prime}, \uparrow}^{\dagger}(\boldsymbol{r}),-\hat{\psi}_{B, \boldsymbol{K}, \downarrow}(\boldsymbol{r}), \hat{\psi}_{A, \boldsymbol{K}, \downarrow}(\boldsymbol{r})\right]^{\dagger}~,
\end{aligned}
}
\end{equation}
where $v_{\rm D} \sim c/300$ is the Fermi velocity in monolayer graphene, the set composed by the two-dimensional matrix identity $\openone_\sigma$ ($\openone_\tau$) and the Pauli matrices $\{ \sigma_x,\sigma_y, \sigma_z \}$ ($\{ \tau_x,\tau_y, \tau_z \}$) act on the two-dimensional sublattice A-B (electron-hole) subspace.
The superconducting order parameter $\Delta(\boldsymbol{r})$, and the scalar potential $U(\boldsymbol{r})$ are defined across the junction as follows
\begin{equation}\label{eq:steplike_parameters}
\begin{aligned}
\Delta(\boldsymbol{r})&=\Theta(|x|-L / 2) \Delta e^{i \phi_{0}(x)}, \\
    \phi_{0}(x)&=\Theta(x) \phi_{\mathrm{R}}+\Theta(-x) \phi_{\mathrm{L}},
\end{aligned}
\end{equation}
\begin{equation}
U(\boldsymbol{r})=-\mu_{0} \Theta(L / 2-|x|)-U_{0} \Theta(|x|-L / 2),
\end{equation}
where $\Theta(x)$ is the Heaviside step function, and we assume that $U_0 \gg |\mu_0|$~\cite{titov_prb_2006}.
The multiple coherent Andreev reflections at the interfaces between the normal phase region and the superconductive sides at $x=\pm L/2$ generate a set of ABSs~\cite{tinkham_book}. These eigenstates of the D-BdG Hamiltonian have subgap eigenenergies, {\rm i.e.} $ |E| <\Delta$. In the short junction limit $L\ll \xi$, where $\xi = \hbar \vD/(\pi \Delta)$ is the coherence length, the eigenstates with energies above the gap, {\rm i.e.} $ |E| >\Delta$, form the Andreev continuum. These states have eigenenergies that are independent of the phase difference $\phi=\phi_{\rm R}-\phi_{\rm L}$, so ABSs are the only ones responsible for carrying the Josephson equilibrium supercurrent~\cite{levchenko_prb_2006,samuelsson_prb_2000}.
In this study, we disregard the continuum and focus on the low-energy characteristics of the GJJ.
To obtain ABSs, we seek for the eigenstates of the D-BdG Hamiltonian by exploiting the uniformity of the system along the $y$-direction
\begin{equation}\label{eq:phi_xyfactorized}
    \boldsymbol{\varphi}_{k, E,\zeta}(\boldsymbol{r})=\frac{e^{i k y}}{\sqrt{W}} \tilde{\boldsymbol{\varphi}}_{k, E}(x),
\end{equation}
where $k$ determines the wavenumber along the $y$ direction, 
and $E$ is the eigenenergy.
Since the D-BdG Hamiltonian in Eq.~\eqref{eq:DBdG_Hamiltonian} is independent of the valley index $\zeta$, we omit it in $\tilde{\boldsymbol{\varphi}}_{k, E}(x)$, and it leads to double degeneracy in the ABS spectrum.
As shown in Fig.~\ref{fig:GJJ}, there are three sharp partitioned sectors along the $x$ direction for which we look for a solution to the stationary D-BdG problem of the following form
\begin{equation}
\tilde{\boldsymbol{\varphi}}_{k, E}(x) = \begin{cases}
     {\cal N}_{k,E} \tilde{\boldsymbol{\varphi}}_{k, E}^{(\mathrm{S}-\mathrm{L})}(x) & x\leq -L/2, \\
    {\cal N}_{k,E} \tilde{\boldsymbol{\varphi}}_{k, E}^{(\mathrm{N})}(x) & -L/2< x< L/2, \\
    {\cal N}_{k,E}  \tilde{\boldsymbol{\varphi}}_{k, E}^{(\mathrm{S}-\mathrm{R})}(x) & x \geq L/2,
   \end{cases}
\end{equation}
and afterward, we impose the continuity at the interfaces $x=\pm L/2$. The coefficient ${\cal N}_{k,E}$ denotes the normalization factor.
To search for eigenfunctions, we follow the approach used in Ref.~\onlinecite{pellegrino_commphys_2022} where the junction length was taken infinitesimal $L/\xi \to 0$. Here, we generalize to the case of a finite junction length $L$.
We start from the superconductive sides ($x<-L/2$ and $x>L/2$).
We write both $ \tilde{\boldsymbol{\varphi}}_{k, E}^{(\mathrm{S}-\mathrm{L})}(x)$ and $ \tilde{\boldsymbol{\varphi}}_{k, E}^{(\mathrm{S}-\mathrm{R})}(x)$ as a linear combination of terms of the form $e^{s \eta_\lambda x} \boldsymbol{a}_{k, E, \lambda,s}$,  where $s=\pm$, $\lambda=\pm$. By imposing the exponential form, it solves the D-BdG equation in the superconductive sides for
\begin{equation}\label{eq:eigenvalues}
     \eta_{\lambda}= i \sqrt{\left(\frac{U_{0}+\lambda \sqrt{E^{2}-\Delta_{0}^{2}}}{\hbar v_{\mathrm{D}}}\right)^{2}-k^{2}}~.
\end{equation}
We exclude pairs of coefficients $\lambda$ and $s$ such that $(\lambda,s)=(+,+)$ and $(-,-)$ (where $\lambda s=+$) on the left-hand side, and $(\lambda,s)=(+,-)$ and $(+,-)$ (where $\lambda s=-$) on the right-hand side, since they correspond to
wavefunctions that cannot be normalized.
Then, we obtain the following expressions
\begin{equation}\label{eq:SCwavefunc_left}
\begin{aligned}
        \boldsymbol{\varphi}_{k, E}^{(\mathrm{S}-\text {L})}(\boldsymbol{r})=&e^{i k y}\left[x^\mathrm{L}_{k,E} e^{-\eta_+(x-L/2)}\boldsymbol{a}_{k, E,+,-}\right. \\
        +&\left.y^\mathrm{L}_{k,E} e^{\eta_-(x-L/2)} \boldsymbol{a}_{k, E,-,+} \right],
\end{aligned}
\end{equation}
\begin{equation}\label{eq:SCwavefunc_right}
\begin{aligned}
        \boldsymbol{\varphi}_{k, E}^{(\mathrm{S}-\text {R})}(\boldsymbol{r})=&e^{i k y}\left[x^\mathrm{R}_{k,E} e^{\eta_+ (x-L/2)}\boldsymbol{a}_{k, E,+,+}\right. \\
        +&\left. y^\mathrm{R}_{k,E} e^{-\eta_- (x-L/2)} \boldsymbol{a}_{k, E,-,-} \right] .
 \end{aligned}
\end{equation}
The four-dimensional vectors $\boldsymbol{a}_{k, E, \lambda, s}$ can be represented as
\begin{equation}\label{eq:eigenvector_w_b}
    \boldsymbol{a}_{k, E, \lambda, s}={\cal W}_{\lambda, s} \boldsymbol{b}_{k, E, \lambda}~,
\end{equation}
where ${\cal W}_{\lambda, s}=\openone_\tau (Q_{\lambda, s} \Lambda_{\lambda, s})$ is defined in terms of the matrices
\begin{equation} \label{diagonalizing_operators}
    \begin{aligned}
    Q_{\lambda, s}&=\frac{1}{\sqrt{2}}\left[\sigma_{z}+i\left(e^{-s z_{\lambda}} \sigma_{-}-e^{s z_{\lambda}} \sigma_{+}\right)\right],\\
        \Lambda_{\lambda, s}&=\frac{1}{\sqrt{2\left[1+e^{s\left(z_{\lambda}+z_{\lambda}^{*}\right)}\right]}}\left(1-\sigma_{z}\right)+\\
    &+\frac{1}{\sqrt{2\left[1+e^{-s\left(z_{\lambda}+z_{\lambda}^{*}\right)}\right]}}\left(1+\sigma_{z}\right),
    \end{aligned}
\end{equation}
which act on the sublattice space, where $z_\lambda$ satisfies the identity $e^{s z_\lambda}=k v_{\rm D}(k+s \eta_\lambda)/[U_0+\lambda \sqrt{E^2-\Delta}]$, and
\begin{equation}\label{eq:eigenvectors_b}
    \boldsymbol{b}_{k, E, \lambda}=\frac{1}{\sqrt{2}}\left[ e^{i \left(\phi_j + \frac{\lambda}{2}\beta(E) \right)}, 0, e^{-i \frac{\lambda}{2} \beta(E)}, 0\right]^{\mathrm{T}},
\end{equation}
with $\beta(E)= \arccos \left(E / \Delta_{0}\right)$ and $j=R,L$.
Assuming that the superconductive sides are in the large doping regime, {\rm i.e.} $U_0 \gg |\mu_0|$, and $\Delta \ll U_0$, then $\eta_\lambda$, defined in Eq.~(\ref{eq:eigenvalues}), can be approximated~\cite{andreev_jetp_1964} as
\begin{equation}\label{eq:AndreevApprox}
    \eta_{\lambda} \approx  i \frac{U_{0}}{\hbar v_{\mathrm{D}}}- \lambda \frac{\sqrt{\Delta_{0}^{2}-E^{2}}}{\hbar v_{\mathrm{D}}}~,
\end{equation}
and $z_\lambda \approx i \pi /2$, which leads to ${\cal W}_{\lambda,s}\approx \frac{1}{\sqrt{2}}\left(\sigma_{z}+s \sigma_{x}\right)$.
To find the solution of the stationary equation in the normal phase sector $\tilde{\boldsymbol{\varphi}}_{k, E}^{(\mathrm{N})}(x)$, we employ the transfer matrix approach~\cite{pellegrino_prb_2011}.
After a few algebraic manipulations, the stationary problem can be rewritten as the following ordinary differential equation in terms of the transfer matrix
\begin{equation}\label{eq:evolution_tranfermatrix}
    \tilde{\boldsymbol{\varphi}}_{k, E}^{(\mathrm{N})}(x)=\mathcal{T}(k, E ; x) \tilde{\boldsymbol{\varphi}}_{k, E}^{(\mathrm{N})}(-L / 2),
\end{equation}
\begin{equation}\label{eq:ODE_normalphase}
    \frac{d {\cal T}(k,E;x)}{d x}=\left[i \tau_{z} \sigma_{x} \frac{E}{\hbar v_{\mathrm{D}}}+i \sigma_{x} \frac{\mu_{0}}{\hbar v_{\mathrm{D}}}+\sigma_{z} k\right] {\cal T}(k,E;x)~,
\end{equation}
with boundary condition $\mathcal{T}(k, E ; x=-L / 2)=\mathbb{1}_{\tau} \mathbb{1}_{\sigma}$.
The transfer matrix which solves the above problem is explicitly expressed as
\begin{equation}
  {\cal T}(k,E;x) = \frac{\openone_\tau+\tau_z}{2} {\mathbb T}(k,E;x) +  \frac{\openone_\tau-\tau_z}{2} {\mathbb T}(k,-E;x)~,
\end{equation}
where
\begin{equation}
\begin{aligned}
{\mathbb T}(k,\pm E;x)&= \sin[q_\pm (x+L/2)] \left[i\frac{\mu_0\pm E}{\hbar \vD q_\pm} \sigma_x+\frac{k}{q_\pm} \sigma_z \right]\\
&+ \cos[q_\pm (x+L/2)]\openone_\sigma~,
\end{aligned}
\end{equation}
and $q_\pm =\sqrt{\left(\frac{\mu_0\pm E}{\hbar \vD } \right)^2 -k^2}$.
Using the transfer matrix, we can impose the continuity condition on the wave functions at the N-S interfaces, and we obtain the compact condition
\begin{equation}\label{eqn:dxTsx}
\tilde{\bm \varphi}^{\rm (S-R)}_{k,E}(x=L/2)={\cal T}(k,E;L/2)\tilde{\bm \varphi}^{\rm (S-L)}_{k,E}(x=-L/2)~.
\end{equation}
After further algebraic manipulations, it can be rewritten as
\begin{equation}\label{eqn:D=0}
\begin{aligned}
&x^{\rm R}_{k,E}   \boldsymbol{a}_{k,E,+,+} +y^{\rm R}_{k,E}   \boldsymbol{a}_{k,E,-,-}= \\
&={\cal T}(k,E;L/2)[x^{\rm L}_{k,E}   \boldsymbol{a}_{k,E,+,-}+y^{\rm L}_{k,E}   \boldsymbol{a}_{k,E,-,+}]~.
\end{aligned}
\end{equation}
The expression above is a homogeneous system of four equations for the variables $\{ x^{\rm L}_{k,E},y^{\rm L}_{k,E},x^{\rm R}_{k,E},y^{\rm R}_{k,E} \}$. The solution of this system is not trivial only if the associated matrix is singular, {\rm i.e.} its determinant $\cal D$ is zero.
For each pair of values of the $y$ component of the wave vector $k$ and the phase difference $\phi$, the energies $E(j,k,\phi)$ that nullify the determinant $\cal D$ are the eigenenergies of the ABSs, with $j$ being the subband index.
The determinant $\cal D$ is an even function of energy, which leads to an even number of subbands $E(j,k,\phi)$, so the subband index runs over the range $\{\pm 1, \pm2,\ldots, \pm n \}$, where $2n$ is the number of subbands, and the following property $E(j,k,\phi)=-E(-j,k,\phi)$ is valid.
For each ABS, after finding the eigenenergy $E(j,k,\phi)$, we choose the nontrivial solution that guarantees the condition $|x^{\rm L}_{k,E(j,k,\phi)}|^2+|y^{\rm L}_{k,E(j,k,\phi)}|^2=1$.

\begin{figure}[t]
\centering
\begin{overpic}[width=\columnwidth]{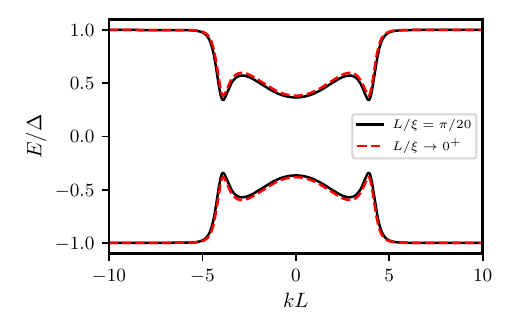}\put(5,55){\normalsize (a)}\end{overpic}\vspace{0em}
\begin{overpic}[width=\columnwidth]{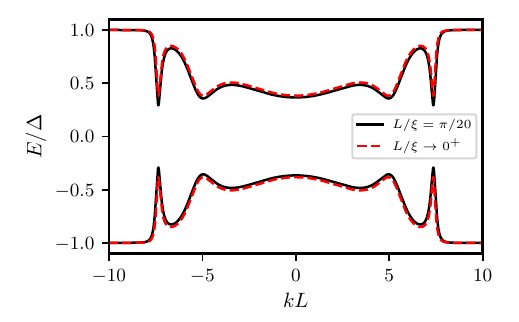}\put(5,55){\normalsize (b)}\end{overpic}\vspace{0em}
\caption{\justifying ABS spectra.
Eigenenergies, in units of the energy gap $\Delta$, as a function of the transverse component of the wavevector $k$, in units of $1/L$.
The ABS spectrum (solid line) for a short junction with $L/\xi=\pi/20$, obtained numerically, is compared to the analytical spectrum (dashed line) in the limit case $L/\xi\to0^+$. Here, we set the Fermi level at $\mu_0=5 \hbar \vD/L$ (a) and $\mu_0=8 \hbar \vD/L$ (b), and the phase difference at the generic value $\phi=\phi_\mathrm{R}-\phi_\mathrm{L}=3 \pi/4$.
\label{fig:ABS}}
\end{figure}
Under the approximation in Eq.~\eqref{eq:AndreevApprox}, the normalization factor ${\cal N}_{k,E(j,k,\phi)}$ is determinated for a given subband index $j$, $y$-component of the wavevector $k$, and phase difference $\phi$, by enforcing the condition below
\begin{equation}
\int^{W/2}_{-W/2} dy \int^{\infty}_{-\infty}d x \boldsymbol{\varphi}^\dagger_{k, E,\zeta}(\boldsymbol{r})  \boldsymbol{\varphi}_{k, E,\zeta}(\boldsymbol{r})\Big|_{E=E(j,k,\phi)} =1~,
\end{equation}
which gives
\begin{equation}
\begin{aligned}
{\cal N}_{k,E(j,k,\phi)} &=  \frac{1}{\sqrt{1+B(j,k, \phi)}} \sqrt{\frac{\sqrt{\Delta^2-E(j,k,\phi)^2}}{\hbar v_{\mathrm{D}} }}~,\\
&
\end{aligned}
\end{equation}
where
\begin{equation}
    \begin{aligned}
       B(j,k, \phi)=&\frac{L}{\xi} \sqrt{1-\left(E(j,k,\phi) / \Delta\right)^{2}}\\
       \times&\left[x_{L} \boldsymbol{a}_{k, E,+,-}+y_{L} \boldsymbol{a}_{k, E,-,+}\right]^{\dagger} \mathcal{A}(k, E(j,k,\phi)) \\
    \times& \left[x_{L} \boldsymbol{a}_{k, E,+,-}+y_{L} \boldsymbol{a}_{k, E,-,+}\right]~,
    \end{aligned}
\end{equation}
\begin{equation}
    \begin{aligned}
    \mathcal{A}(k, E)&=\int_{-L / 2}^{L / 2} \frac{d x}{L} \mathcal{T}^{\dagger}(k, E, x) \mathcal{T}(k, E, x)~\\
    &= \frac{\openone_\tau+\tau_z}{2} {\mathbb A}(k,E) +  \frac{\openone_\tau-\tau_z}{2} {\mathbb A}(k,-E)~,
    \end{aligned}
\end{equation}
\begin{equation}
    \begin{aligned}\label{eq:bbA}
\mathbb{A}(k,\pm E)&=\bigg[\frac{(\mu_0\pm E)^2}{\hbar^2 \vD^2 q^2_\pm }-\frac{k^2\sin(2q_\pm L) }{2q_\pm ^3 L}\bigg]\openone_\sigma  \\
&
-\bigg[\frac{k(\mu_0\pm E)}{\hbar \vD q_\pm }-\frac{k(\mu_0\pm E)\sin(2q_\pm L)}{2\hbar \vD q^3_\pm L}\bigg]\sigma_y~\\
&+
\frac{k \sin^2(q_{\pm}L)}{q^2_\pm L}\sigma_z~,
    \end{aligned}
\end{equation}
and $x^{\rm L}_{k,E}$ and $y^{\rm L}_{k,E}$ are two components of the nontrivial vector which solves the system in Eq.~\eqref{eqn:D=0} with $E\to E(j,k,\phi)$.

From now on, we focus on the short junction regime $L/\xi\ll 1$, where the subband index $j$ can assume two values, namely $\{-1,1\}$, and we have $E(j,k,\phi)=j\epsilon(k,\phi)$.
In particular, we will consider a finite value of $L/\xi\ll 1$, and note that in the limiting case $L/\xi \to 0$, we have $B(j,k, \phi)\to 0$.
To describe the low-energy electronic properties, we project the full D-BdG Hamiltonian in Eq.~\eqref{eq:DBdG_Hamiltonian} onto the ABS subspace with the projector $\hat{\mathcal{P}}_A$~\cite{pellegrino_commphys_2022}.
The Andreev Hamiltonian $\mathcal{H}_A=\hat{\mathcal{P}}_A \hat{\mathcal{H}}_{\mathrm{D}-\mathrm{BdG}}\hat{\mathcal{P}}_A$ is expressed as
\begin{equation}
    \hat{\mathcal{H}}_{\mathrm{A}}=\sum_{\zeta= \pm} \sum_{k} \epsilon(k, \phi) \Big( \hat{\gamma}_{ \zeta,+, k}^{\dagger} \hat{\gamma}_{ \zeta,+, k}-\hat{\gamma}_{ \zeta, -,k}^{\dagger} \hat{\gamma}_{\zeta,-,  k} \Big),
\end{equation}
where  $\hat{\gamma}_{j, \zeta, k}$ is the fermionic ABS operator.
Each set of values of ${\zeta,k}$ determines a two-level system with energy splitting $2\epsilon(k,\phi)$.
Fig.~\ref{fig:ABS} shows the ABS spectrum as a function of the component $k$, with the phase difference set at $\phi=3\pi/4$ and the chemical potential in the normal phase at $\mu_0 = 5 \hbar \vD /L$ (a) and $\mu_0 = 8 \hbar \vD /L$ (b). The solid black line is the spectrum obtained numerically for a finite short junction $L/\xi=\pi/20$, while the red dashed line is the universal analytical expression $\pm {\cal E}(k,\phi)=\pm \Delta \sqrt{1-\tau(k) \sin^2(\phi/2)}$, obtained in the limit $L/\xi\to 0^+$, where the $k$ dependence is totally included in the transmission probability $\tau(k)$ which depends on the nature of the normal phase stripe~\cite{beenakker_book1992}.
For the graphene electron gas~\cite{titov_prb_2006}, the transmission probability of the normal state is $\tau(k)=(\kF^2-k^2)/[\kF^2-k^2 \cos ^2 (L \sqrt{\kF^2-k^2} )]$, with $\kF=\mu_0/(\hbar \vD)$.
From the comparison of the dispersion relations in Fig.~\ref{fig:ABS}, we see that the extrema are located at the same $k$, and in the limiting case $L/\xi\to 0^+$ they correspond to the extrema of the transmission probability~\cite{pellegrino_commphys_2022}.
In Fig.~\ref{fig:ABS} one can see that there is an energy window, which we call mini-gap, in which quasiparticle excitations are prohibited~\cite{banszerus_arxiv_2021}. The minigap $\delta(\phi)=\min_k[\epsilon(k,\phi)]$ depends on the phase difference $\phi$.
In particular, in the limit $L/\xi\to 0^+$, we have $\delta(\phi)= \Delta |\cos(\phi/2)|$.

We conclude this Section by introducing the particle density and the charge density operators, which are the quantities of interest.
Within the D-BdG formalism, the particle density and the charge density operators are expressed, in terms of the four component spinors defined in Eq.~\eqref{eq:spinors}, as
\begin{equation}\label{eq:second_quant_rho}
    \begin{aligned}
        \hat{\rho}(\boldsymbol{r})=&\sum_{\zeta} \hat{\Psi}_{\zeta}^{\dagger}(\boldsymbol{r}) \openone_\tau \openone_\sigma \hat{\Psi}_{\zeta}(\boldsymbol{r}),\\
        \hat{\rho}_{\rm C}(\boldsymbol{r})=&\sum_{\zeta} \hat{\Psi}_{\zeta}^{\dagger}(\boldsymbol{r}) \tau_z \openone_\sigma \hat{\Psi}_{\zeta}(\boldsymbol{r}),
    \end{aligned}
\end{equation}
where in $\hat{\rho}_{\rm C}$ electrons (holes) have charge $+1$ ($-1$).
%
Applying the projector $\hat{\cal P}_A$ on the density operators in Eq.~\eqref{eq:second_quant_rho}, we define the Andreev particle and charge densities operators as
\begin{equation}\label{eq:rhoA}
    \begin{aligned}
        \hat{\rho}_{\rm A}(\boldsymbol{r})=&\hat{\cal P}_A\hat{\rho}(\boldsymbol{r})\hat{\cal P}_A=
        \sum_{\zeta} \sum_{j,j'} \sum_{k,k'} \boldsymbol{\varphi}^\dagger_{k, E(j,k,\phi),\zeta}(\boldsymbol{r})\\
        &\times \openone_\tau \openone_\sigma  \boldsymbol{\varphi}_{k', E(j',k',\phi),\zeta}(\boldsymbol{r})
        \hat{\gamma}_{ \zeta,j, k}^{\dagger} \hat{\gamma}_{ \zeta,j', k'}
        ,\\
        \hat{\rho}_{{\rm C,A}}(\boldsymbol{r})=&\hat{\cal P}_A\hat{\rho}_{\rm C}(\boldsymbol{r})\hat{\cal P}_A=
        \sum_{\zeta} \sum_{j,j'} \sum_{k,k'} \boldsymbol{\varphi}^\dagger_{k, E(j,k,\phi),\zeta}(\boldsymbol{r})\\
        &\times \tau_z \openone_\sigma  \boldsymbol{\varphi}_{k', E(j',k',\phi),\zeta}(\boldsymbol{r})
        \hat{\gamma}_{ \zeta,j, k}^{\dagger} \hat{\gamma}_{ \zeta,j', k'}
        ,
    \end{aligned}
\end{equation}
whose support is the ABS subspace.
In the following sections, we study modifications of the spin-resolved local particle density of states (LDOS).
The spin-up electron and the spin-down hole LDOS are defined as
\begin{equation}\label{eq:LDOS_def}
 \begin{aligned}
     \rho_{{\rm e}\uparrow}(\boldsymbol{r}, \Omega)&={\rm Tr}\left[
     \hat{\rho}_{\rm A, e \uparrow}(\boldsymbol{r})\delta(\Omega-\hat{\cal H}_{\rm tot})\right]~,\\
    \hat{\rho}_{\rm A, e \uparrow}(\boldsymbol{r})&=\frac{1}{2}[\hat{\rho}_{\rm A}(\boldsymbol{r})+\hat{\rho}_{\rm C,A}(\boldsymbol{r})]~,
 \end{aligned}
\end{equation}
\begin{equation}\label{eq:LDOS_defh}
 \begin{aligned}
     \rho_{{\rm h}\downarrow}(\boldsymbol{r}, \Omega)&={\rm Tr}\left[
      \hat{\rho}_{\rm A, h \downarrow}(\boldsymbol{r})\delta(\Omega-\hat{\cal H}_{\rm tot})\right]~,\\
    \hat{\rho}_{\rm A, h \downarrow}(\boldsymbol{r})&=\frac{1}{2}[\hat{\rho}_{\rm A}(\boldsymbol{r})-\hat{\rho}_{\rm C,A}(\boldsymbol{r})]~,
 \end{aligned}
\end{equation}
where we have ${\rm Tr}[\cdot]=\sum_{\zeta,j,k} \langle \emptyset|\hat{\gamma}_{\zeta,j,k} \cdot  \hat{\gamma}^\dagger_{\zeta,j,k}| \emptyset \rangle $, where the reference state $| \emptyset \rangle $ is the vacuum of all field operators $\hat{\gamma}_{\zeta,j,k}$.
The spin-down electron LDOS is given by $\rho_{{\rm h}\downarrow}(\boldsymbol{r}, \Omega)=\rho_{{\rm e}\downarrow}(\boldsymbol{r}, -\Omega)$.

So far we considered the bare GJJ in the short-junction regime.
In the following, we include the presence of a localized impurity described in terms of the Anderson or Lifshitz model.
The total Hamiltonian $\hat{\cal H}_{\rm tot}$ is composed of the Andreev Hamiltonian $\hat{\cal H}_{\rm A}$ and a term $\hat{\cal H}_{\rm imp}$ that takes into account the interaction of the ABSs with a single impurity.

\begin{figure*}[t]
  \centering
  \vspace{1.em}
  \begin{overpic}[width=0.99\columnwidth]{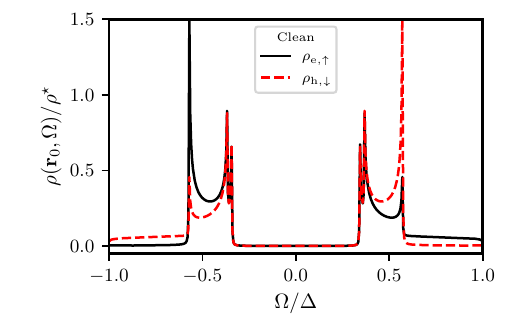}\put(5,55){\normalsize (a)}\end{overpic}
  \begin{overpic}[width=0.99\columnwidth]{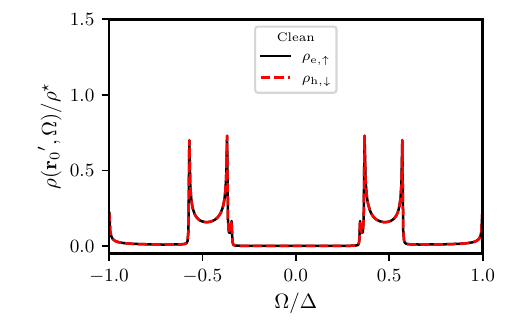}\put(5,55){\normalsize (b)}\end{overpic}\\
  \begin{overpic}[width=0.99\columnwidth]{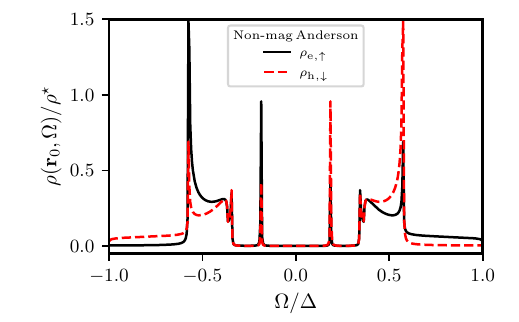}\put(5,55){\normalsize (c)}\end{overpic}
  \begin{overpic}[width=0.99\columnwidth]{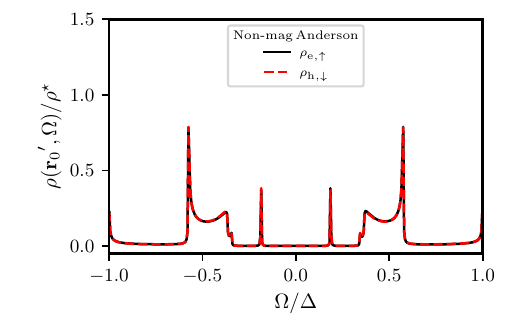}\put(5,55){\normalsize (d)}\end{overpic}\\
  \begin{overpic}[width=0.99\columnwidth]{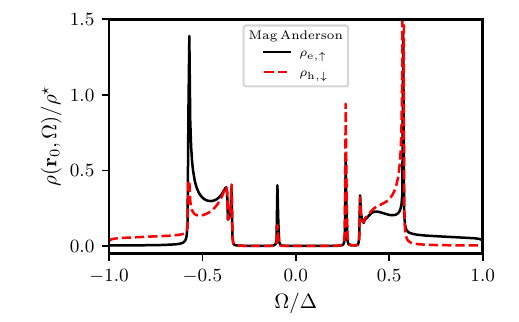}\put(5,55){\normalsize (e)}\end{overpic}
    \begin{overpic}[width=0.99\columnwidth]{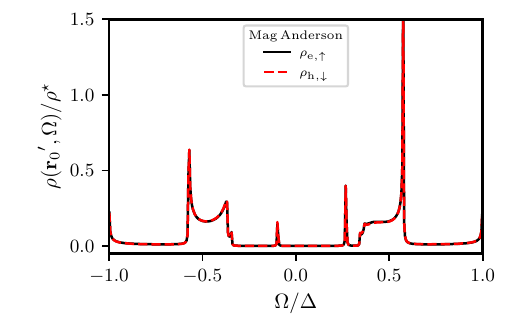}\put(5,55){\normalsize (f)}\end{overpic} \\
  \caption{
\justifying (Color online)  LDOS $\rho_{\rm e,\uparrow}(\bm{r},\Omega)$ (solid black line) and $\rho_{\rm h,\downarrow}(\bm{r},\Omega)$ (red dashed line) as a function of $\Omega/\Delta$, in units of $\rho^\star=1/(\hbar \vD W)$.
Panel (a) and (b) show the LDOS in the clean case calculated in the normal phase region at $\bm{r}_0=(0,0)$ and at the left N-S interface $\bm{r}'_0=(-L/2,0)$, respectively.
In panels (c)  and (d) [(e) and (f)] we consider a single non-magnetic [magnetic] Anderson impurity with  $(\epsilon_{\mathrm{d}},\epsilon_{\mathrm{Z}})=(0.2\Delta,0)$ [$(\epsilon_{\mathrm{d}},\epsilon_{\mathrm{Z}})=(0.2\Delta,0.1\Delta)$] located in the normal phase region at $\bm{r}_d=\bm{r}_0$, and the LDOSs are respectively calculated in $\bm{r}_0$ and $\bm{r}'_0$.
In all panels, the junction length is $L/\xi=\pi/20$, the Fermi level is $\mu_0=5 \hbar \vD/L$,  the tunneling amplitude $t_0=\sqrt{ \Delta \hbar \vD W/(10 A_{\rm c})  }$, the phase difference $\phi=3\pi/4$, and $\eta=10^{-3}\Delta$.
}  \label{fig:LDOS_vs_omega_Anderson}
\end{figure*}

\section{Single Anderson Impurity}
\label{sect:anderson}
In this Section, we discuss the Anderson model~\cite{anderson_physrev_1961,pellegrino_commphys_2022} which describes an impurity that causes inelastic scattering processes in a GJJ electronic system. Within this model, each impurity has two localized electronic states, distinguished by the spin projection along the $z$ direction. For the spin-up and spin-down cases, the energies are $E_{d, \uparrow}=\mu_0+\epsilon_d + \epsilon_{\rm Z}$ and $E_{d,\downarrow}=\mu_0+\epsilon_d - \epsilon_{\rm Z}$, respectively. The
energy $\epsilon_d$ is the impurity energy with respect to the Fermi level $\mu_0$, and the Zeeman energy term $\epsz$  breaks the time-reversal symmetry, and here any Coulomb repulsive term~\cite{capecelatro_prb_2023} is neglected.
Within the Bogoliubov-de Gennes approach, the Hamiltonian of the impurity can be expressed in a spinorial form as
\begin{equation}\label{eq:Himp}
    \begin{aligned}
        \hat{\mathcal{H}}_{\rm D}&= \hat{\Phi}_{d}^{\dagger}\left (\epsilon_{\mathrm{d}} \tau_{z}+\epsilon_{\mathrm{Z}}\openone_{\tau} \right)\hat{\Phi}_{d}
    \end{aligned}
\end{equation}
where $\hat{\Phi}_{d}^{\dagger}=[c_{d,\uparrow}^{\dagger}, c_{d,\downarrow}]$, and the fermionic operators $c_{d,\uparrow}^{\dagger}$ ($ c_{d,\downarrow}$) creates (annihilates) an electron with spin-up (spin-down) bounded to the impurity.
%
Here, we describe the interaction between a single impurity and the electron system in a graphene monolayer in terms of a tunneling Hamiltonian~\cite{pellegrino_commphys_2022} expressed as
\begin{equation}\label{eq:imp_potential}
\begin{aligned}
\hat{\cal V}_{\rm D}&= \hat{\cal V}_{d}+\hat{\cal V}^\dagger_{d}~,\\
    \hat{\cal V}_{d}&=\sum_{\zeta= \pm} \int d^{2} \boldsymbol{r} \hat{\Phi}_{d}^{\dagger} V_{d, \zeta}(\boldsymbol{r}) \hat{\Psi}_{\zeta}(\boldsymbol{r}),
\end{aligned}
\end{equation}
where the matrices $V_{d,\zeta}$ are defined as
\begin{equation}\label{eq:Vd}
    \begin{aligned}
        V_{d,+}(\boldsymbol{r}) & =\left[\begin{array}{cccc}
        v_{A, d}(\boldsymbol{r}) & v_{B, d}(\boldsymbol{r}) & 0 & 0 \\
        0 & 0 & -v_{A, d}^*(\boldsymbol{r}) & -v_{B, d}^*(\boldsymbol{r})
        \end{array}\right], \\
        V_{d,-}(\boldsymbol{r}) & =\left[\begin{array}{cccc}
        -v_{B, d}^*(\boldsymbol{r}) & v_{A, d}^*(\boldsymbol{r}) & 0 & 0 \\
        0 & 0 & v_{B, d}(\boldsymbol{r}) & -v_{A, d}(\boldsymbol{r})
        \end{array}\right].
\end{aligned}
\end{equation}
We assume that the impurity is located at $\bm{r}_d$ in a carbon site and that the tunneling term acts on the electron system in graphene at the atomic scale.
Following Ref.~\onlinecite{pellegrino_commphys_2022}, using a microscopic description based on a tight-binding model for the electron system in graphene, the matrix elements of the potential $\hat{\cal V}_{\rm D}$ are expressed as
\begin{equation}\label{eq:shortrangev}
    \begin{aligned}
        v_{\alpha, d}(\boldsymbol{r})= & t_0 \sqrt{A_c}\big[m_d  \delta_{\alpha, A}  + (1-m_d) \delta_{\alpha, B}\big] \delta\left(\boldsymbol{r}-\boldsymbol{r}_d\right),
    \end{aligned}
\end{equation}
where $A_c=3\sqrt{3}a^2/2$ is the area of the unit cell \cite{katsnelson_book}, with $a=1.42$~\AA, $t_0$ is the tunneling amplitude which, without loss of generality, is taken real and positive. If the impurity is located at a carbon site belonging to the sublattice $A$ ($B$) then one has $m_d=1$ ($m_d=0$).
Here, we consider that the carbon site where the impurity is located belongs to sublattice $A$, and we have $ v_{\alpha, d}(\boldsymbol{r})= t_0 \sqrt{A_{\rm c}}\delta_{\alpha, A}\delta(\boldsymbol{r}-\boldsymbol{r}_d)$.
The results remain unchanged regardless of the sublattice chosen, as a result of the sublattice symmetry.
The local interaction potential projected in the ABS subspace is explicitly written as
\begin{equation}\label{app:onsite_potential}
  \hat{\mathcal{V}}_{d} \hat{\cal P}_{\rm A} = t_0\sum_j \sum_\zeta \sum_k \hat{\Phi}_{d}^{\dagger} w_{d, \zeta,j,k} \hat{\gamma}_{\zeta,j,k}
\end{equation}
where
\begin{equation}\label{eq:w_def}
\begin{aligned}
w_{d,\pm,j,k}&=g_{d,\pm} \boldsymbol{\varphi}_{k, E(j,k,\phi),\pm}  (\boldsymbol{r}_d)\\
&=g_{d,\pm} \frac{e^{i k y_d}}{\sqrt{W}} \tilde{\boldsymbol{\varphi}}_{k, E(j,k,\phi)}(x_d),
\end{aligned}
\end{equation}
\begin{equation}\label{eq:g_def}
\begin{aligned}
g_{d, +}&=  \sqrt{A_c}
  \begin{bmatrix}
1&0&0&0 \\
0&0&-1&0
\end{bmatrix}~,\\
 g_{d, -}&=  \sqrt{A_c}
  \begin{bmatrix}
0&1&0&0 \\
0&0&0&-1
\end{bmatrix} ~.
\end{aligned}
\end{equation}
To study the local effect induced by an Anderson impurity on the electron system in a GJJ, we use the total Hamiltonian $\hat{\mathcal{H}}_{\rm tot}=\hat{\mathcal{H}}_{\rm A}+\hat{\mathcal{H}}_{\rm imp}$, where
$\hat{\mathcal{H}}_{\rm imp}=\hat{\cal H}_{\rm D}+ \hat{\mathcal{V}}_d \hat{\mathcal{P}}_{\mathrm{A}}+ \hat{\mathcal{P}}_{\mathrm{A}} \hat{\mathcal{V}}^{\dagger}_d$, which consists of the impurity Hamiltonian $\hat{\cal H}_{\rm D}$, and the tunneling terms which connect the ABSs and the Anderson impurity.
To calculate the LDOS, we use the following expressions
\begin{equation}\label{eq:LDOS_A}
\begin{aligned}
\rho_{\rm e \uparrow}(\boldsymbol{r}, \Omega)&=-\frac{1}{\pi}{\rm Im} {\rm Tr}\left[  \hat{\rho}_{\rm A,e \uparrow}(\boldsymbol{r}) \hat{\mathcal{G}}_{\rm tot}(\Omega+i\eta)\right],\\
\rho_{\rm h \downarrow}(\boldsymbol{r}, \Omega)&=
-\frac{1}{\pi}{\rm Im} {\rm Tr}\left[  \hat{\rho}_{\rm A,h \downarrow}(\boldsymbol{r}) \hat{\mathcal{G}}_{\rm tot}(\Omega+i\eta)\right],\\
\end{aligned}
\end{equation}
which involve the total Green function
\begin{equation}
\hat{\mathcal{G}}_{\rm tot}(\Omega)=(\Omega \openone-\hat{\mathcal{H}}_{\rm tot})^{-1}~,
\end{equation}
with $\eta=0^+$.
After some algebraic manipulations, we write the projection of the total Green function onto the ABS subspace as
\begin{equation}\label{eq:Greensfunction_effectiveH}
\begin{aligned}
\hat{\mathcal{G}}(\Omega)=&\hat{\mathcal{P}}_{\mathrm{A}} \hat{\mathcal{G}}_{\rm tot}(\Omega) \hat{\mathcal{P}}_{\mathrm{A}}=\left(\Omega\openone-\hat{\mathcal{H}}_{\mathrm{eff}}\right)^{-1},\\
\hat{\mathcal{H}}_{\mathrm{eff}}=&\hat{\mathcal{H}}_{\mathrm{A}}+\hat{\mathcal{P}}_{\mathrm{A}} \hat{\mathcal{V}}_{d}^{\dagger}\left(\Omega \openone-\hat{\mathcal{H}}_{d}\right)^{-1} \hat{\mathcal{V}}_{d} \hat{\mathcal{P}}_{\mathrm{A}}.
\end{aligned}
\end{equation}

Before calculating the LDOS in the presence of a single Anderson impurity, we consider the clean GJJ.
For the unperturbed GJJ system, the ABS Green function takes the simple form
\begin{equation}\label{eq:G0}
\begin{aligned}
\hat{\mathcal{G}}_{0}(\Omega)&=\left(\Omega\openone-\hat{\mathcal{H}}_{\mathrm{A}}\right)^{-1}  \\
&=\sum_{j=\pm} \sum_{\zeta= \pm} \sum_{k} \frac{1}{\Omega-j \epsilon( k, \phi)}\hat{\gamma}_{j, \zeta, k}^{\dagger} \hat{\gamma}_{j, \zeta, k},
 \end{aligned}
\end{equation}
and by replacing it in Eq.~\eqref{eq:LDOS_A}, we find the LDOS
\begin{equation}\label{eq:rho0}
\begin{aligned}
&\rho_{0,\ell }(\boldsymbol{r}, \Omega)=-\frac{1}{\pi}{\rm Im} {\rm Tr}\left[\hat{\rho}_{{\rm A},\ell}(\boldsymbol{r}) \hat{\mathcal{G}}_{0}(\Omega+i\eta )\right].
\end{aligned}
\end{equation}
%
%
It is useful to introduce the following $4\times4$ matrix Green function
\begin{equation}\label{eq:Gbar}
\begin{aligned}
&\bar{G}_0(\boldsymbol{r},\boldsymbol{r}',\Omega)=\sum_j \sum_k
\frac{\boldsymbol{\varphi}_{k, E(j,k,\phi),\zeta} (\boldsymbol{r}) \boldsymbol{\varphi}^\dagger_{k, E(j,k,\phi),\zeta} (\boldsymbol{r}')}{\Omega -j\epsilon(k,\phi)} \\
&=  \sum_k  \frac{e^{ik(y-y')}}{W} \Bigg[ \sum_j \frac{\tilde{\bm \varphi}_{k, E(j,k,\phi)}(x) \tilde{\bm \varphi}^\dagger_{k, E(j,k,\phi)}(x')}{\Omega -j\epsilon(k,\phi)} \Bigg],
\end{aligned}
\end{equation}
which acts on the direct product of the electron-hole subspace and sub-lattice subspace on which the spinors of the D-BdG formalism are built. It allows one to write a compact and intuitive form of the LDOSs.
Although the valley index $\zeta$ is indicated in the initial line, the spinorial wavefunction does not depend on it, so it is omitted in $\bar{G}_0(\boldsymbol{r},\boldsymbol{r}',\Omega)$.
Furthermore, looking at the second line of Eq.~\eqref{eq:Gbar}, each entry of $\bar{G}_0(\boldsymbol{r},\boldsymbol{r}',\Omega)$ has the form of a Fourier transform, and we calculate them using the fast Fourier transform algorithm~\cite{kong_python_2020}, see details in Appendix~\ref{app:nFT}.
Starting from the definition in Eq.~\eqref{eq:rho0}, applying the cyclic property of the trace, and using Eq.~\eqref{eq:Gbar}, the LDOS in the clear limit takes the following compact form
\begin{equation}\label{eq:rho0-compact}
\rho_{0,\ell }(\boldsymbol{r}, \Omega)=-\frac{1}{\pi}
{\rm Im} \left\{ {\rm tr}[M_\ell \bar{G}_0(\boldsymbol{r},\boldsymbol{r},\Omega+i \eta)] \right\},
\end{equation}
where $\ell=\{{\rm e \uparrow},{\rm h \downarrow} \}$, $M_{\rm e \uparrow}=\frac{1}{2}(\openone_\tau + \tau_z) \openone_\sigma$, and $M_{\rm h \downarrow}=\frac{1}{2}(\openone_\tau - \tau_z) \openone_\sigma$, and ${\rm tr}[\cdot]$ denotes the trace over the direct product of the electron-hole subspace and sub-lattice subspace.

To calculate the LDOS in the presence of a single impurity, one has to find the ABS Green function $ \hat{\mathcal{G}}(\Omega)$ expressed in Eq.~\eqref{eq:Greensfunction_effectiveH}.
To this aim, we rewrite Eq.~\eqref{eq:Greensfunction_effectiveH} in the form of a Dyson equation as
\begin{equation}\label{eq:Dyson}
 \hat{\mathcal{G}}(\Omega)=\hat{\mathcal{G}}_{0}(\Omega)+\hat{\mathcal{G}}_{0}(\Omega)  \hat{\Sigma}(\Omega) \hat{\mathcal{G}}(\Omega)~,
\end{equation}
where the self-energy is
\begin{equation}\label{eq:DysonA}
\begin{aligned}
 \hat{\Sigma}(\Omega)&=  \hat{\mathcal{P}}_{\mathrm{A}} \hat{\mathcal{V}}_{d}^{\dagger}\left(\Omega \openone-\hat{\mathcal{H}}_{\rm D}\right)^{-1} \hat{\mathcal{V}}_{d} \hat{\mathcal{P}}_{\mathrm{A}}.
\end{aligned}
\end{equation}
The exact solution to the Dyson equation, as detailed in Appendix~\ref{app:ABS-G}, is expressed in the following closed form
\begin{equation}\label{eq:G_F}
    \hat{\mathcal{G}}(\Omega)  =\hat{\mathcal{G}}_{0}(\Omega)+\sum_{\zeta,\zeta'}\sum_{j, j^{\prime}} \sum_{k, k^{\prime}} F_{\zeta,j,k,\zeta',j',k'} (\Omega)\hat{\gamma}_{\zeta,j,  k}^{\dagger} \hat{\gamma}_{\zeta^{\prime}, j^{\prime}, k^{\prime}},
\end{equation}
where the analytical expression of $ F_{\zeta,j,k,\zeta',j',k'}(\Omega)$ is reported in Appendix~\ref{app:ABS-G}, see Eq.\eqref{app:F}.
Based on the results of Eq.\eqref{eq:G_F}, for both electrons and holes, the LDOS is written as the sum of two terms $\rho_\ell(\boldsymbol{r}, \Omega)=\rho_{\ell,0}(\boldsymbol{r}, \Omega)+\delta \rho_\ell(\boldsymbol{r}, \Omega)$, where
%
\begin{equation}\label{eq:drho_def}
\begin{aligned}
&\delta \rho_{\ell }(\boldsymbol{r}, \Omega)=-\frac{1}{\pi}{\rm Im} {\rm Tr}\left\{\hat{\rho}_{{\rm A},\ell}(\boldsymbol{r})[\hat{\mathcal{G}}(\Omega+i\eta )- \hat{\mathcal{G}}_{0}(\Omega+i\eta )]\right\}.
\end{aligned}
\end{equation}
Using Eq.~\eqref{eq:G_F}, we have the following
\begin{equation}\label{eq:drho+F}
\begin{aligned}
&\delta \rho_{\ell }(\boldsymbol{r}, \Omega)=-\frac{1}{\pi}{\rm Im} \sum_{\zeta}
\sum_{j,j'} \sum_{k,k'} \boldsymbol{\varphi}^\dagger_{k, E(j,k,\phi),\zeta}(\boldsymbol{r}) M_\ell \\
&\times \boldsymbol{\varphi}_{k', E(j',k',\phi),\zeta}(\boldsymbol{r}) F_{\zeta,j',k',\zeta,j,k}(\Omega+i \eta)~,
\end{aligned}
\end{equation}
where $\eta=0^+$.
Applying the cyclic property of the trace and using Eqs.~\eqref{eq:g_def} and \eqref{eq:Gbar}, the expression above can be rewritten as
\begin{widetext}
\begin{equation}\label{eq:drho}
\begin{aligned}
&\delta \rho_{\ell }(\boldsymbol{r}, \Omega)=-\frac{1}{\pi}
{\rm Im} {\rm tr} \Big\{M_\ell \bar{G}_0(\bm{r},\bm{r}_d,\Omega+i\eta)
\sum_\zeta g_{d,\zeta}^\dagger X(\Omega + i \eta) [\openone_\tau - Y(\Omega+i\eta)]^{-1} g_{d,\zeta} \bar{G}_0(\bm{r}_d,\bm{r},\Omega+i\eta)  \Big\}~,
\end{aligned}
\end{equation}
\end{widetext}
where
\begin{equation}\label{eq:Y-G}
Y(\Omega)= \sum_\zeta g_{d,\zeta}\bar{G}_0({\bm r}_d,{\bm r}_d,\Omega) g^\dagger_{d,\zeta}X(\Omega)~,
\end{equation}
and the nature of the impurity is included in the $2\times2$ matrix $X(\Omega)$. For an Anderson impurity, one has
\begin{equation}\label{eq:X-Anderson}
X(\Omega) \to t_0^2 G_d(\Omega)~,
\end{equation}
with
\begin{equation}\label{eq:Gd_def}
\begin{aligned}
G_d(\Omega) &=  \left[(\Omega-\epsilon_{\mathrm{Z}})\openone_{\tau} -\epsilon_{\mathrm{d}} \tau_{z}\right]^{-1}\\
&=
\begin{bmatrix}
\frac{1}{\Omega-(\epsilon_d+\epsilon_{\rm Z})}&0 \\
0&\frac{1}{\Omega+(\epsilon_d-\epsilon_{\rm Z})}
\end{bmatrix}~.
\end{aligned}
\end{equation}

Fig.~\ref{fig:LDOS_vs_omega_Anderson} shows the LDOS for electrons with spin-up (solid black line) and holes with spin-down (red dashed line) as a function of energy $\Omega$. The junction length, Fermi level and phase difference are set to $L/\xi=\pi/20$, $\mu_0=5 \hbar \vD/L$, and $\phi=3\pi/4$, respectively.
To obtain the LDOS numerically, a finite value of $\eta=10^{-3}\Delta$ is used, which produces a fictitious finite linewidth of the energy levels within the mini-gap and a smoothing of the square root divergences that are expected at the extrema of the dispersion relations $E(j,k)$.
Figs.~\ref{fig:LDOS_vs_omega_Anderson}~(a) and (b) illustrate the case of clean GJJ, evaluated in the normal phase region at $\bm{r}_0=(0,0)$ and at the N-S interface $\bm{r}_0'=(-L/2,0)$, respectively. In both panels, the LDOS $\rho_{0,{\rm e \uparrow}}$ and $\rho_{0,{\rm h \downarrow}}$ are zero for energies within $ |\Omega| < \delta(\phi)$.
Figs.~\ref{fig:LDOS_vs_omega_Anderson}~(c) and (d) illustrate the LDOS in $\bm{r}_0=(0,0)$ and $\bm{r}'_0=(-L/2,0)$, respectively, when a non-magnetic Anderson impurity with parameters $(\epsilon_d,\epsilon_{\rm Z})=(0.2,0) \Delta$ is present at $\bm{r}_d=\bm{r}_0$. Figs.~\ref{fig:LDOS_vs_omega_Anderson}~(e) and (f) show the same LDOS when the impurity is magnetic, with parameters $(\epsilon_d,\epsilon_{\rm Z})=(0.2,0.1) \Delta$.
For both Anderson impurities, we fix the tunneling amplitude at $t_0=\sqrt{ \Delta \hbar \vD W/(10 A_{\rm c})  }$.
In the absence of any spin-splitting term, such as the Zeeman term, the total Hamiltonian $\hat{\cal H}_{\rm tot}$ is particle-hole symmetric. This symmetry leads to the identity $\rho_{\rm e,\uparrow}(\bm{r},\Omega)=\rho_{\rm h,\downarrow}(\bm{r},-\Omega)$. The relation between the spin-down electron and the spin-down hole LDOS, which is independent of the Hamiltonian, is $\rho_{\rm e,\uparrow}(\bm{r},\Omega)=\rho_{\rm e,\downarrow}(\bm{r},\Omega)$. Therefore, with a particle-hole symmetric Hamiltonian, the electron LDOS is independent of the spin $z$-projection, {\rm i.e.} $\rho_{\rm e,\uparrow}(\bm{r},\Omega)=\rho_{\rm e,\downarrow}(\bm{r},\Omega)$.
In the presence of a non-magnetic Anderson impurity where the particle-hole symmetry is present (see Fig.~\ref{fig:LDOS_vs_omega_Anderson} (c) and (d)), two bound states within the mini-gap, due to the hybridization of ABS and impurity, arise at opposite energies, with $\Omega_0=-0.185 \Delta \gtrsim -\epsilon_d$ and $\Omega_0'=-\Omega_0$.
In Fig.~\ref{fig:LDOS_vs_omega_Anderson}~(e) and (f), the effect of a magnetic Anderson impurity is taken into account, which breaks the particle-hole symmetry. This is evidenced by the emergence of two sharp peaks within the minigap at energies $\Omega_0=-0.098 \Delta \gtrsim -\epsilon_d + \epsilon_{\rm Z}$ and $\Omega_{0}'=0.267 \Delta \lesssim \epsilon_d + \epsilon_{\rm Z}$, which are no longer opposite in energy, $\Omega_0'\neq -\Omega_0$.
\begin{figure*}[t]
\centering
\begin{overpic}[width=0.99\columnwidth]{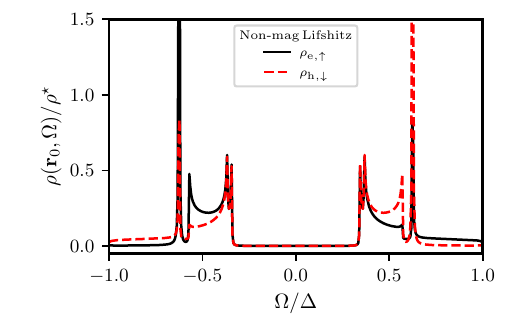}\put(5,55){\normalsize (a)}\end{overpic}
    \begin{overpic}[width=0.99\columnwidth]{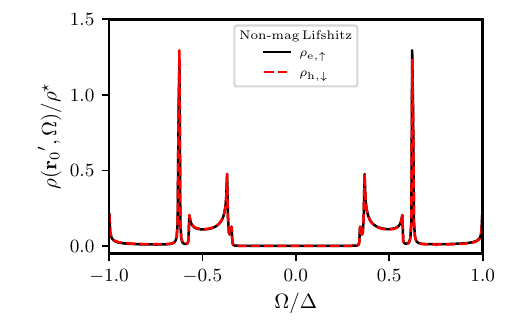}\put(5,55){\normalsize (b)}\end{overpic} \\
  \begin{overpic}[width=0.99\columnwidth]{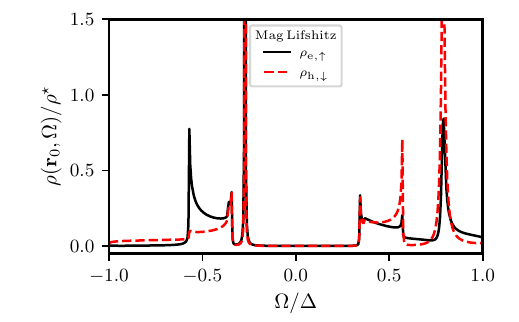}\put(5,55){\normalsize (c)}\end{overpic}
    \begin{overpic}[width=0.99\columnwidth]{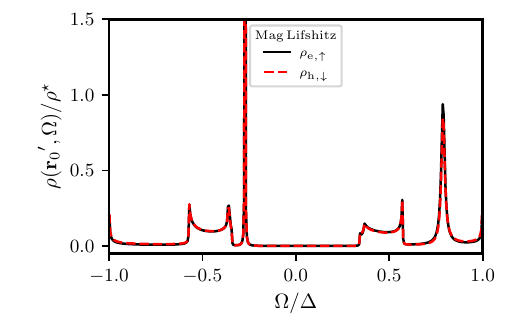}\put(5,55){\normalsize (d)}\end{overpic} \\
\caption{
\justifying   (Color online)  LDOS $\rho_{\rm e,\uparrow}(\bm{r},\Omega)$ (solid black line) and $\rho_{\rm h,\downarrow}(\bm{r},\Omega)$ (red dashed line) as a function of $\Omega/\Delta$, in units of $\rho^\star=1/(\hbar \vD W)$.
In all panels, the junction length is $L/\xi=\pi/20$, the Fermi level is $\mu_0=5 \hbar \vD/L$,  the phase difference $\phi=3\pi/4$, and $\eta=10^{-3}\Delta$.
In panels (a) and (b)  [(c) and (d)] the LDOS are respectively calculated in the normal phase region at $\bm{r}_0=(0,0)$ and at the left N-S interface $\bm{r}'_0=(-L/2,0)$,  and the GJJ is affected by a single non-magnetic [magnetic] Lifshitz impurity with $(u_{\mathrm{d}},u_{\mathrm{Z}})=(-1/8,0)  \hbar \vD W/A_{\rm c}$ [$(u_{\mathrm{d}},u_{\mathrm{Z}})=(-1/8,0.1225) \hbar \vD W/A_{\rm c}$]. }
\label{fig:LDOS_vs_omega_Lifshitz}
\end{figure*}

\begin{figure*}[t]
\centering
\begin{overpic}[width=0.99\columnwidth]{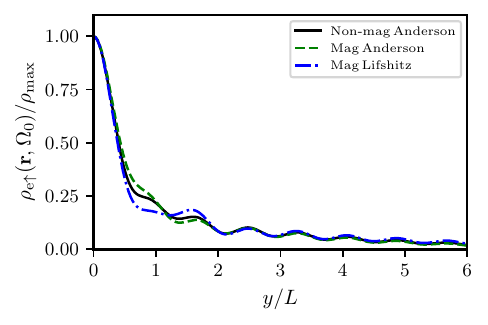}\put(2,60){\normalsize (a)}\end{overpic}
    \begin{overpic}[width=0.99\columnwidth]{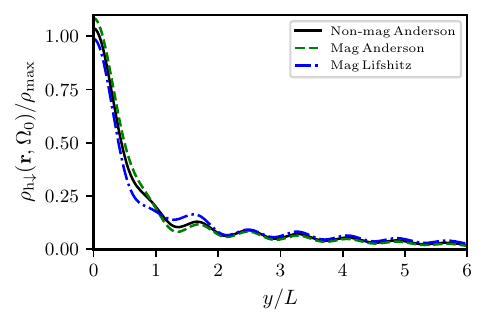}\put(2,60){\normalsize (b)}\end{overpic} \\
  \begin{overpic}[width=0.99\columnwidth]{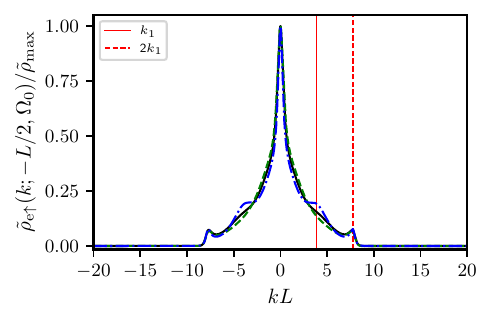}\put(2,60){\normalsize (c)}\end{overpic}
    \begin{overpic}[width=0.99\columnwidth]{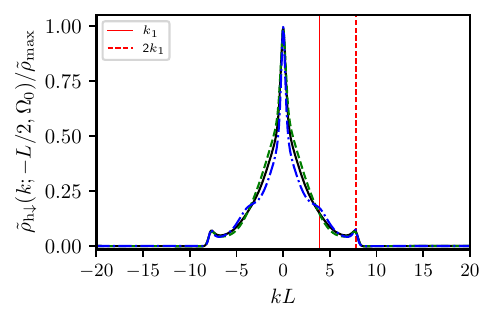}\put(2,60){\normalsize (d)}\end{overpic}
\caption{ \justifying (Color online)
LDOS of electrons with spin-up $\rho_{\rm e \uparrow}(\bm{r},\Omega_0)$ (a) and holes with spin-down $\rho_{\rm h \downarrow}(\bm{r},\Omega_0)$ (b), respectively, evaluated as a function of the coordinate $y/L$, along the N-S interface $\bm{r}=(-L/2,y)$. We consider a single impurity located in the normal phase region at $\bm{r}_d=(0,0)$, and the energy value $\Omega_0$ corresponds to a sharp peak within the mini-gap induced by the impurity. The electron and hole LDOSs are expressed in units of $\rho_{\rm max}=\rho_{\rm e \uparrow}((-L/2,0)),\Omega_0)$.
The solid black line and the green dashed line refer to an Anderson impurity: the first is non-magnetic, with $(\epsilon_{\mathrm{d}},\epsilon_{\mathrm{Z}})=(0.2\Delta,0)$ and $\Omega_0=-0.185 \Delta$, and the second is magnetic, with $(\epsilon_{\mathrm{d}},\epsilon_{\mathrm{Z}})=(0.2\Delta,0.1\Delta)$  and $\Omega_0=-0.098 \Delta$, the tunneling amplitude $t_0$ for both cases is equal to $\sqrt{\Delta \hbar \vD W/(10 A_{\rm c})}$.
The dashed-dotted blue line refers to a magnetic Lifshitz impurity with $(u_{\mathrm{d}},u_{\mathrm{Z}})=(-1.8,0.1225) \hbar \vD W/A_{\rm c}$ and $\Omega_0=-0.272 \Delta$.
Panels (c) and (d) show the partial Fourier transforms $\tilde{\rho}_{\rm e \uparrow}(k;-L/2, \Omega_0)$ and $\tilde{\rho}_{\rm h \downarrow}(k;-L/2, \Omega_0)$, defined in Eq.~\eqref{eq:partial_FT} as a function of $kL$, calculated, respectively, for the cases shown in (a) and (b), and in units of $\tilde{\rho}_{\rm max}=\tilde{\rho}_{\rm e \uparrow}(0;-L/2,\Omega_0)$.
 Here, the vertical lines indicate $k_1$ (red solid line) and $2k_1$ (red dashed line), where $k_1 L=\sqrt{(\mu_0 L)^2/(\hbar \vD)^2-\pi^2}$.
In all panels,  the junction length is $L/\xi=\pi/20$, the Fermi level is $\mu_0=5 \hbar \vD/L$,  the phase difference $\phi=3\pi/4$, and $\eta=10^{-3}\Delta$.
}
\label{fig:LDOS_space_SN}
\end{figure*}
\begin{figure*}[t]
\centering
\begin{overpic}[width=0.99\columnwidth]{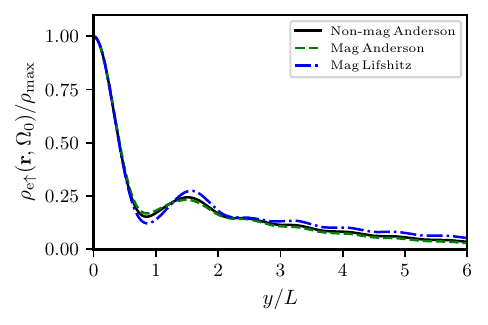}\put(2,60){\normalsize (a)}\end{overpic}
    \begin{overpic}[width=0.99\columnwidth]{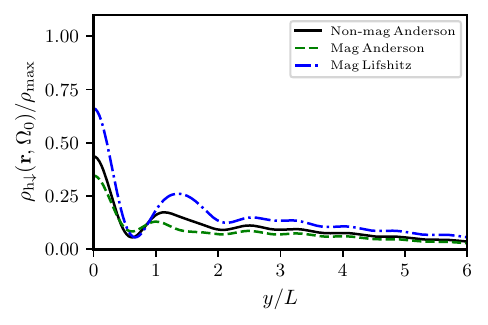}\put(2,60){\normalsize (b)}\end{overpic} \\
  \begin{overpic}[width=0.99\columnwidth]{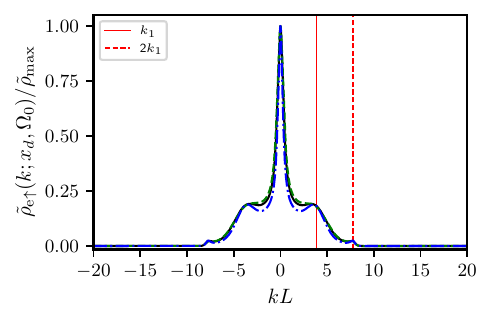}\put(2,60){\normalsize (c)}\end{overpic}
    \begin{overpic}[width=0.99\columnwidth]{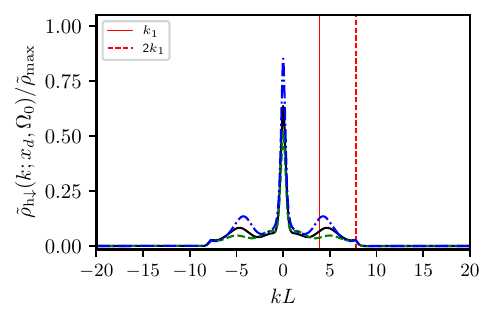}\put(2,60){\normalsize (d)}\end{overpic}
\caption{ \justifying (Color online)
LDOS of electrons with spin-up $\rho_{\rm e \uparrow}(\bm{r},\Omega_0)$ (a) and holes with spin-down $\rho_{\rm h \downarrow}(\bm{r},\Omega_0)$ (b),  respectively, evaluated as a function of the coordinate $y/L$, where $\bm{r}=(x_d,y)$. We consider a single impurity located in the normal phase region at $\bm{r}_d=(0,0)$, and the energy value $\Omega_0$ corresponds to a sharp peak within the mini-gap induced by the impurity. The LDOSs of both electrons and holes are expressed in units of $\rho_{\rm max}=\rho_{\rm e \uparrow}(\bm{r}_d,\Omega_0)$.
The solid black line and the green dashed line refer to an Anderson impurity: the first is non-magnetic, with $(\epsilon_{\mathrm{d}},\epsilon_{\mathrm{Z}})=(0.2\Delta,0)$ and $\Omega_0=-0.185 \Delta$, and the second is magnetic, with $(\epsilon_{\mathrm{d}},\epsilon_{\mathrm{Z}})=(0.2\Delta,0.1\Delta)$  and $\Omega_0=-0.098 \Delta$, the tunneling amplitude for both cases is equal to $\sqrt{\Delta \hbar \vD W/(10 A_{\rm c})}$.
The dashed-dotted blue line refers to a magnetic Lifshitz impurity with $(u_{\mathrm{d}},u_{\mathrm{Z}})=(-1.8,0.1225) \hbar \vD W/A_{\rm c}$ and $\Omega_0=-0.272 \Delta$.
Panels (c) and (d) show the partial Fourier transforms $\tilde{\rho}_{\rm e \uparrow}(k;x_d, \Omega_0)$ and $\tilde{\rho}_{\rm h \downarrow}(k;x_d, \Omega_0)$, defined in Eq.~\eqref{eq:partial_FT} as a function of $kL$, calculated respectively for the cases shown in (a) and (b), and in units of $\tilde{\rho}_{\rm max}=\tilde{\rho}_{\rm e \uparrow}(0;x_d,\Omega_0)$.
 Here, the vertical lines indicate $k_1$ (red solid line) and $2k_1$ (red dashed line), where $k_1 L=\sqrt{(\mu_0 L)^2/(\hbar \vD)^2-\pi^2}$.
In all panels,  the junction length is $L/\xi=\pi/20$, the Fermi level is $\mu_0=5 \hbar \vD/L$,  the phase difference is $\phi=3\pi/4$, and $\eta=10^{-3}\Delta$.
}
\label{fig:LDOS_space}
\end{figure*}
\begin{figure*}[t]
\centering
\begin{overpic}[width=0.99\columnwidth]{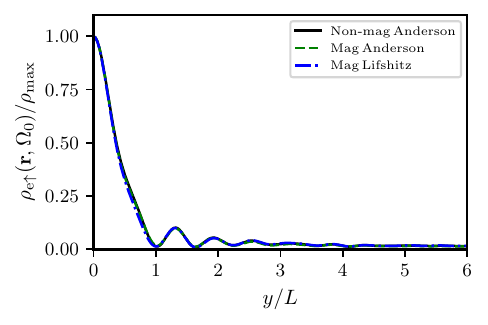}\put(2,60){\normalsize (a)}\end{overpic}
    \begin{overpic}[width=0.99\columnwidth]{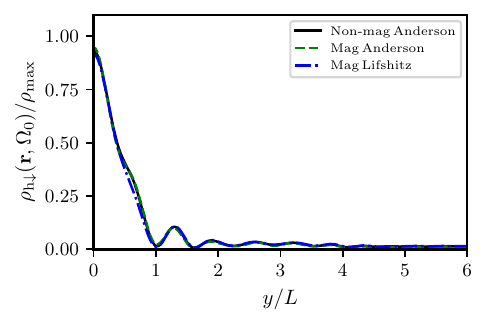}\put(2,60){\normalsize (b)}\end{overpic} \\
  \begin{overpic}[width=0.99\columnwidth]{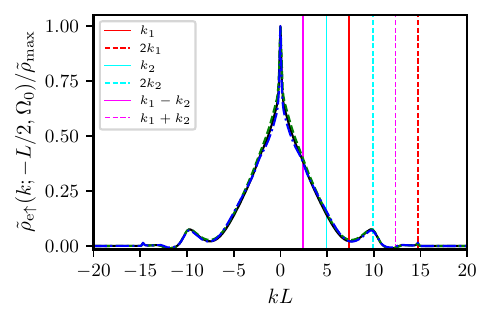}\put(2,60){\normalsize (c)}\end{overpic}
    \begin{overpic}[width=0.99\columnwidth]{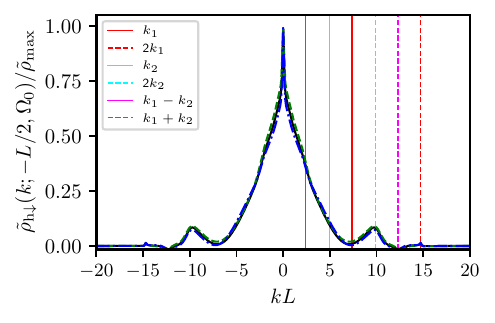}\put(2,60){\normalsize (d)}\end{overpic} \\
\caption{ \justifying (Color online)
LDOS of electrons with spin-up (a) and holes with spin-down (b), $\rho_{\rm e \uparrow}(\bm{r},\Omega_0)$ and $\rho_{\rm h \downarrow}(\bm{r},\Omega_0)$, respectively, evaluated as a function of the coordinate $y/L$, along the N-S interface $\bm{r}=(-L/2,y)$. We consider a single impurity located in the normal phase region at $\bm{r}_d=(0,0)$, and the energy value $\Omega_0$ corresponds to a sharp peak within the mini-gap induced by the impurity. The LDOSs of both electrons and holes are expressed in units of $\rho_{\rm max}=\rho_{\rm e \uparrow}((-L/2,0),\Omega_0)$.
Solid black line and the green dashed line refer to an Anderson impurity: the first is non-magnetic, with $(\epsilon_{\mathrm{d}},\epsilon_{\mathrm{Z}})=(0.2\Delta,0)$ and $\Omega_0=-0.17 \Delta$, and the second is magnetic, with $(\epsilon_{\mathrm{d}},\epsilon_{\mathrm{Z}})=(0.2\Delta,0.1\Delta)$  and $\Omega_0=-0.09 \Delta$, the tunneling amplitude for both cases is equal to $\sqrt{\Delta \hbar \vD W/(10 A_{\rm c})}$.
The dashed-dotted blue line refers to a magnetic Lifshitz impurity with $(u_{\mathrm{d}},u_{\mathrm{Z}})=(-1.8,0.1225) \hbar \vD W/A_{\rm c}$ and $\Omega_0=-0.21 \Delta$.
Panels (c) and (d) show the partial Fourier transforms $\tilde{\rho}_{\rm e \uparrow}(k;-L/2, \Omega_0)$ and $\tilde{\rho}_{\rm h \downarrow}(k;-L/2, \Omega_0)$, defined in Eq.~\eqref{eq:partial_FT} as a function of $kL$, calculated respectively for the cases shown in (a) and (b), and in units of $\tilde{\rho}_{\rm max}=\tilde{\rho}_{\rm e \uparrow}(0;-L/2,\Omega_0)$.
Here, the vertical lines indicate $k_1$ (red solid line), $2k_1$ (red dashed line), $2 k_2$ (cyan solid line), $k_2$ (cyan dashed line), $k_1-k_2$ (magenta solid line), $k_1+k_2$ (magenta dashed line), where $k_n L=\sqrt{(\mu_0 L)^2/(\hbar \vD)^2-n^2 \pi^2}$.
In all panels,  the junction length is $L/\xi=\pi/20$, the Fermi level is $\mu_0=8 \hbar \vD/L$,  the phase difference $\phi=3\pi/4$, and $\eta=10^{-3}\Delta$.
}
\label{fig:LDOS_space_v2_SN}
\end{figure*}
\begin{figure*}[t]
\centering
\begin{overpic}[width=0.99\columnwidth]{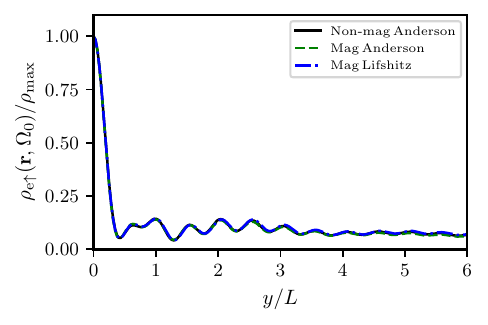}\put(2,60){\normalsize (a)}\end{overpic}
    \begin{overpic}[width=0.99\columnwidth]{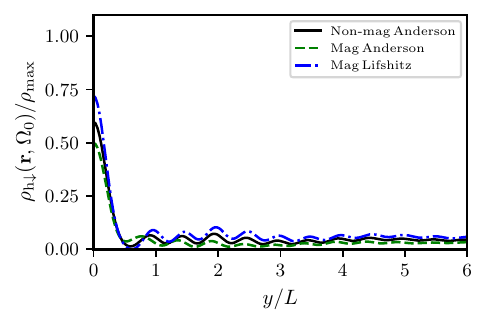}\put(2,60){\normalsize (b)}\end{overpic} \\
  \begin{overpic}[width=0.99\columnwidth]{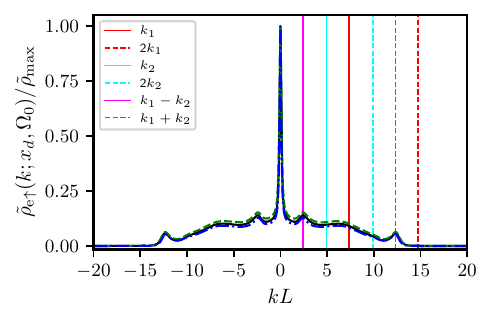}\put(2,60){\normalsize (c)}\end{overpic}
    \begin{overpic}[width=0.99\columnwidth]{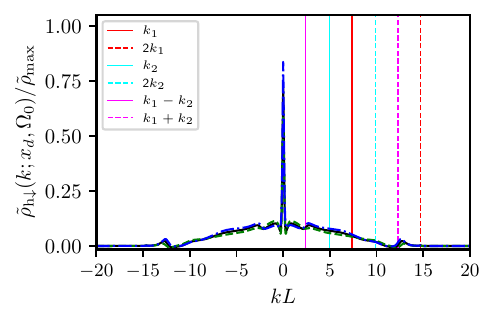}\put(2,60){\normalsize (d)}\end{overpic} \\
\caption{ \justifying (Color online)
LDOS of electrons with spin-up (a) and holes with spin-down (b), $\rho_{\rm e \uparrow}(\bm{r},\Omega_0)$ and $\rho_{\rm h \downarrow}(\bm{r},\Omega_0)$, respectively, evaluated as a function of the coordinate $y/L$, where $\bm{r}=(x_d,y)$. We consider a single impurity located in the normal phase region at $\bm{r}_d=(0,0)$, and the energy value $\Omega_0$ corresponds to a sharp peak within the mini-gap induced by the impurity. The LDOS of both electrons and holes are expressed in units of $\rho_{\rm max}=\rho_{\rm e \uparrow}(\bm{r}_d,\Omega_0)$.
Solid black line and the green dashed line refer to an Anderson impurity: the first is non-magnetic, with $(\epsilon_{\mathrm{d}},\epsilon_{\mathrm{Z}})=(0.2\Delta,0)$ and $\Omega_0=-0.17 \Delta$, and the second is magnetic, with $(\epsilon_{\mathrm{d}},\epsilon_{\mathrm{Z}})=(0.2\Delta,0.1\Delta)$  and $\Omega_0=-0.09 \Delta$, the tunneling amplitude for both cases is equal to $\sqrt{\Delta \hbar \vD W/(10 A_{\rm c})}$.
The dashed-dotted blue line refers to a magnetic Lifshitz impurity with $(u_{\mathrm{d}},u_{\mathrm{Z}})=(-1.8,0.1225) \hbar \vD W/A_{\rm c}$ and $\Omega_0=-0.21 \Delta$.
Panels (c) and d) show partial Fourier transforms $\tilde{\rho}_{\rm e \uparrow}(k;x_d, \Omega_0)$ and $\tilde{\rho}_{\rm h \downarrow}(k;x_d, \Omega_0)$, defined in Eq.~\eqref{eq:partial_FT} as a function of $kL$, calculated respectively for the cases shown in (a) and (b), and in units of $\tilde{\rho}_{\rm max}=\tilde{\rho}_{\rm e \uparrow}(0;x_d,\Omega_0)$.
Here, the vertical lines indicate $k_1$ (red solid line), $2k_1$ (red dashed line), $2 k_2$ (cyan solid line), $k_2$ (cyan dashed line), $k_1-k_2$ (magenta solid line), $k_1+k_2$ (magenta dashed line), where $k_n L=\sqrt{(\mu_0 L)^2/(\hbar \vD)^2-n^2 \pi^2}$.
In all panels,  the junction length is $L/\xi=\pi/20$, the Fermi level is $\mu_0=8 \hbar \vD/L$,  the phase difference $\phi=3\pi/4$, and $\eta=10^{-3}\Delta$.
}
\label{fig:LDOS_space_v2}
\end{figure*}

\section{Single Lifshitz impurity}
\label{sect:lifshitz}
In this Section, we consider an impurity that acts as a source of elastic scattering processes for the electron system in a GJJ, and it is modeled by the Lifshitz model~\cite{skrypnyk_prb_2011}.
Similarly to Sect.~\ref{sect:anderson}, we consider a localized impurity that is placed at position $\bm{r}_d$ and at a carbon site that belongs to sublattice $A$, which generates the potential
\begin{equation}
\begin{aligned}
 \hat{\cal U}_{\rm D} &=
\sum_{\boldsymbol{\kappa}'=\boldsymbol{K},\boldsymbol{K}'}
\sum_{\boldsymbol{\kappa}=\boldsymbol{K},\boldsymbol{K}'}
 \sum_{\alpha=A,B}  \sum_{\alpha'=A,B}
  \sum_{s=\uparrow,\downarrow} \int d^2 \bm{r} u_{s}  \\
&\times \delta_{s,s'} \delta_{\alpha,A}\delta_{\alpha',A} A_{\rm c} \delta(\bm{r}-\bm{r}_d)
 \hat{\psi}_{\alpha,\boldsymbol{\kappa}, s}^{\dagger}(\boldsymbol{r})  \hat{\psi}_{\alpha', \boldsymbol{\kappa}', s'}(\boldsymbol{r})~.
\end{aligned}
\end{equation}
Here, $u_s$ is the spin-resolved interaction energy, but spin-flip processes are excluded.
Due to the short-range potential of the impurity, there is a mixing between the valleys $\boldsymbol{K}$ and $\boldsymbol{K}'$.
We focus on the low-energy electronic system and project the total Hamiltonian onto the ABS subspace. The ABS Green function can be expressed in the form of a Dyson equation, see Eq.~\eqref{eq:Dyson}, where the self-energy is given by
\begin{equation}\label{eq:DysonL}
\begin{aligned}
\hat{\Sigma}(\Omega)&=  {\cal P}_{\rm A}\hat{\cal U}_{\rm D}{\cal P}_{\rm A}\\
 &=\sum_{\zeta,\zeta'} \sum_{j,j'} \sum_{k,k'}w^\dagger_{d,\zeta,j,k} U_d w_{d,\zeta',j',k'} \hat{\gamma}_{\zeta,j,  k}^{\dagger} \hat{\gamma}_{\zeta^{\prime}, j^{\prime}, k^{\prime}}~,
\end{aligned}
\end{equation}
with
\begin{equation}\label{eqn:Uddef}
U_d=u_d \tau_z + u_{\rm Z} \openone_\tau~,
\end{equation}
$u_d=(u_{\uparrow}+u_{\downarrow})/2$ and $u_{\rm Z}=(u_{\uparrow}-u_{\downarrow})/2$.
Following the procedure used for a single Anderson impurity (details in Appendix~\ref{app:ABS-G}), we find that the ABS Green function $\hat{\cal G}(\Omega)$ has the form shown in Eq.~\eqref{eq:G_F}, with
%
\begin{equation}\label{eq:X-Lifshitz}
X(\Omega) \to U_d.
\end{equation}
Moreover, for both electrons and holes, the LDOS can be expressed as $\rho_\ell(\boldsymbol{r}, \Omega)=\rho_{\ell,0}(\boldsymbol{r}, \Omega)+\delta \rho_\ell(\boldsymbol{r}, \Omega)$, where $\rho_{\ell,0}(\boldsymbol{r}, \Omega)$ is equal to Eq.~\eqref{eq:rho0-compact}, while the modification induced by a single Lifshitz impurity is given by Eq.~\eqref{eq:drho} with the substitution in Eq.~\eqref{eq:X-Lifshitz}.
Fig.~\ref{fig:LDOS_vs_omega_Lifshitz} shows the LDOS for electrons with spin-up (solid black line) and holes with spin-down (red dashed line) as a function of energy $\Omega$, by selecting the junction length, Fermi level and phase difference used in Fig. ~\ref{fig:LDOS_vs_omega_Anderson}.
Fig.~\ref{fig:LDOS_vs_omega_Lifshitz}~(a) [(c)] and (b) [(d)] show the LDOS at $\bm{r}_0=(0,0)$ and $\bm{r}'_0=(-L/2,0)$, respectively, in the presence of a non-magnetic [magnetic] Lifshitz impurity located at $\bm{r}_d=\bm{r}_0$ and characterized by $(u_{\mathrm{d}},u_{\mathrm{Z}})=(-1/8,0)  \hbar \vD W/A_{\rm c}$ [$(u_{\mathrm{d}},u_{\mathrm{Z}})=(-1/8,0.1225) \hbar \vD W/A_{\rm c}$].
A non-magnetic Lifshitz impurity acts as a source of elastic scattering that does not disrupt Cooper pairs~\cite{anderson_jpcs_1959}. Therefore, for any value of $u_d$ and $u_{\rm z}=0$, the mini-gap is not affected by impurity scattering.
Indeed, in Fig.~\ref{fig:LDOS_vs_omega_Lifshitz}~(a) and (b), the LDOS $\rho_{0,{\rm e \uparrow}}$ and $\rho_{0,{\rm h \downarrow}}$ are exactly zero within the energy range $ \Omega < \delta(\phi)$, despite the presence of a strongly attractive non-magnetic Lifshitz impurity.
In Appendix~\ref{app:noBoundState}, we verify semi-analitically that the emergence of a bound state within the mini-gap energy generated by the non-magnetic Lifshitz impurity is impossible.
The spin-independent strong attractive interaction considered produces a pair of symmetric peaks, $\Omega_0=-0.623 \Delta$ and $\Omega_0'=-\Omega_0$, which appear outside the mini-gap energy window, $\delta(\phi)<|\Omega_0|<\Delta$, and preserves the particle-hole symmetry property $\rho_{\rm e \uparrow}(\bm{r},\Omega)=\rho_{\rm h \downarrow}(\bm{r},-\Omega)$.
In contrast, a magnetic Lifshitz impurity,  which breaks the particle-hole symmetry, is a source of elastic scattering processes that can disrupt Cooper pairs~\cite{balatsky_rmp_2006}. In Fig.~\ref{fig:LDOS_vs_omega_Lifshitz}~(c) and (d), we consider a spin-resolved potential, with $u_\uparrow=u_d+u_{\rm Z}=0.25 \times 10^{-2} \hbar \vD W/A_{\rm c}$ and $u_\downarrow=u_d-u_{\rm Z}=-0.2475 \hbar \vD W/A_{\rm c}$. This term $u_{\rm Z}$ reduces the interaction with the spin-up channel and increases the attractive interaction with the spin-down channel. Two sharp peaks appear at $\Omega_{0}=-0.272 \Delta$ and $\Omega_{0}'=0.786 \Delta$, and only the first one is within the mini-gap energy window.
Moreover, in Fig.~\ref{fig:LDOS_vs_omega_Anderson} and Fig.~\ref{fig:LDOS_vs_omega_Lifshitz}, it is shown that the profiles $\rho_{0,{\rm e \uparrow}}(\bm{r}_0',\Omega)$ and $\rho_{0,{\rm h \uparrow}}(\bm{r}_0',\Omega)$ calculated along the N-S boundary are indistinguishable. This holds independently of the presence or type of impurity (see the right-hand panels). This symmetry emerges from the form of the eigenfunction on the superconducting side, as expressed in Eq.~\eqref{eq:eigenvectors_b}, where the spin-up electron and spin-down hole are equal in magnitude regardless of energy. However, this symmetry is not maintained in the normal-phase region, where the junction has a finite length $L$. In fact, for any generic $k$ and finite energy $E$, the transfer matrix $\mathcal{T}(k, E; x)$ acts differently on the spin-up electron and spin-down hole sectors.

\section{Analysis in real space}
\label{sect:realspace}

In this Section, we investigate the structure of the density fluctuations that are generated near the impurity, which can result in a sharp peak in the LDOS at an energy within the mini-gap energy $\delta(\phi)$.
In the presence of a single impurity at $\bm{r}_d=(x_d,y_d)=(0,0)$, Figs.~\ref{fig:LDOS_space_SN}~(a)-(b) and Figs.~\ref{fig:LDOS_space}~(a)-(b) [Figs.~ \ref{fig:LDOS_space_v2_SN}~(a)-(b) and Figs.~\ref{fig:LDOS_space_v2}~(a)-(b)] show the LDOS for spin-up electrons and spin-down holes, respectively, along the transverse direction $\bm{r}=(x_d,y)$ and the N-S interface $\bm{r}=(-L/2,y)$, with the Fermi level set at $\mu_0=5 \hbar \vD/L$ [$\mu_0=8 \hbar \vD/L$].
In all panels, $\Omega_0$ is the energy at which a sharp peak of the LDOS appears in the range of $-\delta(\phi)<\Omega_0<0$. We have considered an Anderson non-magnetic impurity (solid black line), an Anderson magnetic impurity (green dashed line), and a Lifshitz magnetic impurity (blue dashed-dotted line).
For all of the cases considered, the finite signals of LDOSs shown here are consequences of the hybridization between the impurity and the ABSs. In fact, the LDOSs for a clean GJJ both $\rho_{\rm e \uparrow,0}(\bm{r},\Omega_0)$ and  $\rho_{\rm h \downarrow,0}(\bm{r},\Omega_0)$  are exactly zero.
Moreover, regardless of the nature of the impurity, the LDOS exhibits a similar pattern of oscillations. This oscillatory behavior reminds us of Friedel oscillations that arise in an electron gas from localized perturbations~\cite{Fetter_book}, where their characteristic wavevector $2 \kF$ is a direct consequence of an important property of the perturbed system, namely the existence of a sharp Fermi surface in an electron gas~\cite{giulianivignale_book}.
We now explore the Fourier analysis of the spatial pattern of the LDOS to determine what information about the ABSs in a GJJ can be obtained.
To this aim, we define the partial Fourier transform as
\begin{equation}\label{eq:partial_FT}
    \tilde{\rho}_\ell(k;x,\Omega)=
   \int_{-\infty}^{\infty} d y e^{-i k y} \rho_\ell\left(\bm{r},\Omega \right)~,
\end{equation}
where $\ell=\{{\rm e \uparrow},{\rm h \downarrow}\}$, and they have been calculated by using the fast Fourier transform algorithm~\cite{kong_python_2020} (details in Appendix~\ref{app:nFT}).
Fig.~\ref{fig:LDOS_space_SN}~(c)-(d) and Fig.~\ref{fig:LDOS_space}~(c)-(d)  [Fig.~\ref{fig:LDOS_space_v2_SN}~(c)-(d) and Fig.~\ref{fig:LDOS_space_v2}~(c)  -(d)]  show respectively the partial Fourier transform $\tilde{\rho}_{\ell}(k;-L/2,\Omega_0)$ and $\tilde{\rho}_{\ell}(k;x_d,\Omega_0)$ ($\ell=\{{\rm e \uparrow},{\rm h \downarrow}\}$) of the LDOS illustrated in  Fig.~\ref{fig:LDOS_space_SN}~(a)-(b) and Fig.~\ref{fig:LDOS_space}~(a)-(b)   [Fig.~\ref{fig:LDOS_space_v2_SN}~(a)-(b) and Fig.~\ref{fig:LDOS_space_v2}~(a)  -(b)], and the Fermi level is $\mu_0=5 \hbar \vD/L$ [$\mu_0=8 \hbar \vD/L$].
Independently of the type of impurity and doping level, $\tilde{\rho}_{\ell}(k;-L/2,\Omega_0)$ and $\tilde{\rho}_{\ell}(k;x_d,\Omega_0)$ have a global maximum at $k=0$, indicating that the most effective scattering processes that lead to the bound state at $\Omega_0$ have zero transferred momentum.
In addition to the peak at $k=0$, for both the doping levels under considerations, $\tilde{\rho}_{\ell}(k;-L/2,\Omega_0)$ and $\tilde{\rho}_{\ell}(k;x_d,\Omega_0)$ show a pronunciated profile around finite wavenumbers.
To understand the origin of these special transferred momenta, we recall that Fig.~\ref{fig:ABS} shows how the ABS dispersion relation $\epsilon(k,\phi)$ and the analytical expression ${\cal E}(k,\phi)= \Delta \sqrt{1-\tau(k) \sin^2(\phi/2)}$ have extrema located at the same momenta~\cite{beenakker_book1992}.
In particular, for a given phase difference $\phi$,  the minima of ${\cal E}(k,\phi)$ occur when the transmission probability $\tau(k)$ reaches its maximum value $\tau(k)=1$ (total transmission).
For the normal state of the graphene electron gas~\cite{titov_prb_2006}, the total transmission occurs for $k_0=0$ (Klein tunneling~\cite{katsnelson_book})  and for $\pm k_n=\pm \sqrt{\mu_0^2/(\hbar \vD)^2-(n \pi)^2/L^2}$ (stationary wave condition), where $n=1,\ldots,\lfloor \mu_0L /(\pi \hbar \vD)\rfloor$ ($\lfloor \cdot\rfloor$ is the integer part).
Klein tunneling is independent of the doping level, whereas the number of momenta that satisfy the stationary wave condition is a function of the doping level.
For example, at a doping level $\mu_0=5\hbar \vD/L$, only $\pm k_1$ satisfy the stationary wave condition. When the doping level increases to $\mu_0=8\hbar \vD/L$, $\pm k_1$ and $\pm k_2$ satisfy the stationary wave condition.
At special momenta $k_0$ and $\pm k_n$, which correspond to highly transmissive channels of the normal graphene stripe, the ABS dispersion relation $\epsilon(k,\phi)$ shows quasi-degenerate minima, and at these energy values, the density of states exhibits square root divergences~\cite{pellegrino_commphys_2022}.
At the doping level $\mu_0=5\hbar \vD/L$, regardless of the nature of the impurity, along the N-S interface $x=-L/2$, we observe in Fig.~\ref{fig:LDOS_space_SN}~(a)-(b) an oscillating behavior of spin-up electrons and spin-down holes LDOSs with a characteristic wavenumber $2 k_1$, as confirmed by the partial Fourier transform in  Fig.~\ref{fig:LDOS_space_SN}~(c)-(d), where pronounced peaks around $\pm 2 k_1$ appear.
Along the $x=x_d$ line within the normal phase region, both the spin-up electron and spin-down hole LDOSs show a pattern in Fig.~\ref{fig:LDOS_space}~(a)-(b) that is reminiscent of beats. In fact, in the corresponding partial Fourier transforms in Fig.~\ref{fig:LDOS_space}~(c)-(d) pronounced peaks appear around two wavenumbers $k_1$ and $2 k_1$.
This result suggests that the scattering processes with exchanged momentum $2 k_1$ and $k_1$ play a crucial role in the generation of bound states, and that the ABSs involved have momenta around the special ones $k_0$ and $\pm k_1$.
At the doping level of $\mu_0=8\hbar \vD/L$, the spin-up electron and spin-down hole LDOSs along the N-S interface at $x=-L/2$ display an oscillating behavior with a wavenumber of $2 k_2$, as seen in Fig.~\ref{fig:LDOS_space_v2_SN}~(a)-(b). The partial Fourier transforms in Fig.~\ref{fig:LDOS_space_v2_SN}~(c)-(d) provide further confirmation, with two distinct peaks at $\pm 2 k_2$.
Along the $x=x_d$ line in the normal phase region, both the spin-up electron and spin-down hole LDOSs in Fig.~\ref{fig:LDOS_space_v2_SN}~(a)-(b) show a pattern that resembles beats, and the partial Fourier transforms in Fig.~\ref{fig:LDOS_space_v2_SN}~(c)-(d) are prominent around $\pm(k_1-k_2)$ and $\pm(k_1+k_2)$.
This result indicates that the momenta of ABS more involved in the formation of the bound states are close to the specific values $k_0$, $\pm k_1$, and $\pm k_2$.
In conclusion, the Fourier analysis of LDOS reveals the momenta mainly involved in the scattering processes with ABSs associated with a high transmission probability $\tau(k)$, corresponding to the minima of $\epsilon(k,\phi)$.

\section{Conclusions}
\label{sect:conclusions}

In this work, we investigated the LDOS in the normal phase stripe of a ballistic short GJJ. In particular, we studied spatial effects in the vicinity of a short-range impurity and the generation of bound states resulting from the hybridization with ABSs.
We examined two distinct descriptions of a single short-range impurity, namely the Anderson model and the Lifshitz model. In the Anderson model the impurity is a source of inelastic scattering, whereas in the Lifshitz model scattering is elastic.
We have found that in the absence of magnetic terms, the Lifshitz impurity does not produce any bound state within the mini-gap. This result can be seen as a consequence of the Anderson theorem, which states that this kind of impurity cannot disrupt Cooper pairs. On the other hand, the Anderson impurity is capable of forming two bound states within the minigap, which are symmetric in energy with respect to the Fermi level.
By introducing a magnetic term, we find that the Lifshitz impurity disrupts Cooper pairs, and it can induce a bound state within the mini-gap. For the Anderson model, an additional magnetic term can break the energy symmetry of the two bound states within the mini-gap.
Therefore, from the energy features of the LDOS within the mini-gap, the nature of an impurity coupled to the electron system of a short ballistic GJJ can be distinguished.
We have also studied the spatial structure of the density fluctuations at a bound state within the minigap. We found that, except for the Lifshitz non-magnetic model, for fixed phase difference and Fermi level, the LDOS exhibits a unique oscillating and decreasing trend as one moves away from the impurity, regardless of the nature of the impurity.
For each spatial profile, the corresponding Fourier analysis shows that the characteristic wavevectors of the density oscillations are connected to the momenta of the high-transmissive channels in ballistic graphene, providing insight into the properties of normal-phase graphene.

\begin{acknowledgments}
The authors thank J. Ankerhold, F. Bonasera, P. Hakonen, C. Padurariu, and V. Varrica for illuminating discussions and fruitful comments on various stages of this work. E.P. acknowledges support from the PNRR MUR project PE0000023-NQSTI  and COST Action CA21144 superqumap. F.M.D.P acknowledges support from the project PRIN 2022 - 2022XK5CPX (PE3) SoS-QuBa - "Solid State Quantum Batteries: Characterization and Optimization" and the PNRR MUR project PE0000023-NQSTI.
G.G.N.A. acknowledges support from the project MUR PRIN PNRR - P20223LXTA - ENTANGLE.
G.F. thanks for the support  ICSC - Centro Nazionale di Ricerca in High-Performance Computing, Big Data and Quantum Computing under project E63C22001000006 and Universit\`a degli Studi di Catania, Piano di Incentivi per la Ricerca di Ateneo, project TCMQI.
I.V. thanks the Institut f\"ur Komplexe Quantensysteme and IQST, Universit\"at Ulm (Germany) for the great hospitality and stimulating environment. I.V. and F.M.D.P. contributed equally to this
work.
\end{acknowledgments}

\appendix
\counterwithin{figure}{section}

\section{Calculation of the ABS's Green function with a single impurity}
\label{app:ABS-G}
In this Appendix, we show the details for the calculation of the ABS's Green function in the presence of a short-range impurity.
The approach described below is used for both the Anderson model (see Sect.~\ref{sect:anderson}) and the Lifshitz model (see Sect.~\ref{sect:lifshitz}).
In both cases, the ABS's Green function is involved in a Dyson series, see Eqs.~\eqref{eq:DysonA}-\eqref{eq:DysonL}, of the form
\begin{equation*}
\hat{\cal G}(\Omega)=\hat{\cal G}_0(\Omega)+\hat{\cal G}_0(\Omega) \hat{ \Sigma}(\Omega)\hat{\cal G}(\Omega)~,
\end{equation*}
where $\hat{\cal G}_0(\Omega)$ is defined in Eq.~\eqref{eq:G0} and
\begin{equation}
\hat{ \Sigma}(\Omega)=\sum_{\zeta,\zeta'} \sum_{j,j'} \sum_{k,k'}w^\dagger_{d,\zeta,j,k} X(\Omega) w_{d,\zeta',j',k'} \hat{\gamma}_{\zeta,j,  k}^{\dagger} \hat{\gamma}_{\zeta^{\prime}, j^{\prime}, k^{\prime}}~.
\end{equation}
For the Anderson model, we have the following correspondence
\begin{equation}\label{app:X-Anderson}
X(\Omega) \to
\begin{bmatrix}
\frac{t_0^2}{\Omega-(\epsilon_d+\epsilon_{\rm Z})}&0 \\
0&\frac{t_0^2}{\Omega+(\epsilon_d-\epsilon_{\rm Z})}
\end{bmatrix}~,
\end{equation}
which is equal $t_0^2 G_d(\Omega)$, defined in Eq.~\eqref{eq:Gd_def}. Whereas, for the Lifshitz model, we have
\begin{equation}\label{app:X-Lifshitz}
X(\Omega) \to
\begin{bmatrix}
u_d+u_z&0 \\
0&-u_d+u_z
\end{bmatrix}~,
\end{equation}
which corresponds with $U_d$, defined in Eq.~\eqref{eqn:Uddef}.
By expanding the ABS's Green function as
\begin{equation}\label{app:GF_expansion}
    \hat{\mathcal{G}}(\Omega)=\sum_{n=0}^{\infty}\hat{\mathcal{G}}_n (\Omega),
\end{equation}
where the $n$-th term
\begin{equation}\label{app:series_term}
    \hat{\mathcal{G}}_{n}(\Omega)=\hat{\mathcal{G}}_{0}(\Omega)\left[\hat{ \Sigma}(\Omega) \hat{\mathcal{G}}_{0}(\Omega)\right]^{n}~,
\end{equation}
for $n>0$ it can be expressed as
\begin{widetext}
\begin{eqnarray}\label{app:Gn_termseries}
  \hat{\cal G}_n(\Omega)&=&
 \sum_{j,j^\prime} \sum_{\zeta,\zeta^\prime}
 \sum_{k,k^\prime}
\frac{1}{\Omega-j \epsilon(k,\phi)}
  w^\dagger_{d, j,\zeta,k}
X(\Omega)
Y^{n-1}(\Omega)  w_{d,j',\zeta',k'}  G_0(j^\prime,k^\prime,\Omega)  \hat{\gamma}^\dagger_{j,\zeta,k} \hat{\gamma}_{j^\prime,\zeta^\prime,k^\prime}\nonumber\\
\end{eqnarray}
\end{widetext}
where
\begin{equation}\label{app:Y}
Y(\Omega)  =  \sum_{\zeta,j,k}
 \frac{ w_{d,\zeta,j,k}    w^\dagger_{d,\zeta,j,k}}{\Omega-j \epsilon(k,\phi)}     X(\Omega)~.
\end{equation}
%
%
%
It is convenient to write the ABS's Green function as
\begin{equation}\label{app:Gn_termseries-2}
\begin{aligned}
  \hat{\cal G}(\Omega)&=\hat{\cal G}_0(\Omega) +\sum_{j,j^\prime} \sum_{\zeta,\zeta^\prime}
 \sum_{k,k^\prime}
\frac{1}{\Omega-j \epsilon(k,\phi)}w^\dagger_{d,\zeta,j,k}
X(\Omega) \nonumber \\
  &\times
 \left[ \sum^\infty_{n=0} Y^n(\Omega)\right]  w_{d,\zeta',j',k'}
\frac{1}{\Omega-j' \epsilon(k',\phi)}\hat{\gamma}^\dagger_{j,\zeta,k} \hat{\gamma}_{j^\prime,\zeta^\prime,k^\prime},
\end{aligned}
\end{equation}
in the square brackets, we have isolated the geometric series of ratio $Y(\Omega)$, which can be expressed in a closed form, and $\hat{\mathcal{G}}(\Omega)$ can be written in the form
\begin{equation}\label{app:G_summed}
    \hat{\mathcal{G}}(\Omega)  =\hat{\mathcal{G}}_{0}(\Omega)+\sum_{\zeta,\zeta'}\sum_{j, j^{\prime}} \sum_{k, k^{\prime}} F_{j,\zeta,k,j',\zeta',k'} (\Omega)\hat{\gamma}_{j, \zeta, k}^{\dagger} \hat{\gamma}_{j^{\prime}, \zeta^{\prime}, k^{\prime}},
\end{equation}
where
\begin{equation}\label{app:F}
\begin{aligned}
&   F_{\zeta,j,k,\zeta',j',k'} (\Omega) =\frac{1}{\Omega-j \epsilon(k,\phi)} w^\dagger_{d,\zeta,j,k}X(\Omega)\\
    &\times \left[\openone_\tau-Y(\Omega)\right]^{-1} w_{d,\zeta',j',k'}\frac{1}{\Omega-j' \epsilon(k',\phi)}.
\end{aligned}
\end{equation}
The result expressed in Eq.~\eqref{app:G_summed} is formally valid for both types of impurities, the differences from the Anderson model and the Lifshitz model are generated by the form of the corresponding matrix $X(\Omega)$, see Eqs.~\eqref{app:X-Anderson} and \eqref{app:X-Lifshitz}.

\section{Numerical Fourier Transform}
\label{app:nFT}

The Discrete Fourier Transform (DFT) is a mathematical operation that converts a finite sequence of discrete complex values $\{ z_n \}$ into another sequence $\{ \tilde{z}_j \}$, according to the definition
\begin{equation}\label{app:DFT}
\tilde{z}_j = \sum^{N-1}_{n=0} e^{-i 2\pi j n/N} z_n,
\end{equation}
and the inverse operation (IDFT) is defined as
\begin{equation}\label{app:IDFT}
z_n = \frac{1}{N}\sum^{N-1}_{j=0} e^{i 2\pi j n/N}  \tilde{z}_j.
\end{equation}
Both of these can be calculated rapidly using the Fast Fourier Transform (FFT)~\cite{kong_python_2020}.
We demonstrate how IDFT has been used to calculate the $4\times4$ matrix $\bar{G}_0(\boldsymbol{r},\boldsymbol{r}',\Omega)$. According to Eq.~\eqref{eq:Gbar}, each entry of $\bar{G}_0(\boldsymbol{r},\boldsymbol{r}',\Omega)$ is an inverse Fourier transform from the $k$ domain to the $y$ domain, {\rm i.e.}
\begin{equation}
f(y)= \sum_k e^{i k y} \tilde{f}(k)~.
\end{equation}
In the case of a wide GJJ ($W \gg L,\xi$), the summation over $k$ can be replaced by an integration
\begin{equation}\label{app:IFT}
f(y)= \frac{ W }{2\pi} \int^{\infty}_{-\infty} dk e^{i k y} \tilde{f}(k)~.
\end{equation}
Now, we take $\tilde{f}(k)$ in the range $[-k_{\rm max},k_{\max}]$ and select the finite sequence $\tilde{f}_j=f(k_j-k_{\rm max})$, with $k_j = 2 k_{\rm max} j/N$ and $j=0,\ldots,N-1$. Using the definition of IDFT, we have the following
\begin{equation}
f_n=\frac{1}{N} \sum^{N-1}_{j=0} e^{ iy_n k_j} \tilde{f}(k_j-k_{\rm max})~.
\end{equation}
where $y_n=\pi n /k_{\rm max} $.
So, the resolution in the $y$-domain is $\Delta y = \pi /k_{\rm max} $ and the spatial range is $[-\pi N/(2 k_{\rm max}),\pi N/(2 k_{\rm max})]$.
When $k_{\rm max}$ is kept constant, the discrete values obtained by the IDFT approach the following expression in the limit of large $N\gg1$:
\begin{equation}\label{app:iDFT_cont}
f_n= \frac{1}{2 k_{\rm max}}\int^{k_{\rm max}}_{-k_{\rm max}} d k e^{i k y_n}\tilde{f}(k).
\end{equation}
Comparing with Eq.~\eqref{app:IFT}, we can see that
\begin{equation}
f(y_n)= \lim_{k_{\rm max} \to \infty} \lim_{N \to \infty} \frac{ k_{\rm max} W}{\pi } f_n~.
\end{equation}
To calculate $\bar{G}_0(\boldsymbol{r},\boldsymbol{r}',\Omega)$, we have taken the expression above with the finite values of $N=23062$ and $k_{\rm max}=100/L$.
According to this approximation, the function $f(y)$ is calculated over the spatial range $[-y_{\rm max}, y_{\rm max}]$ where $y_{\rm max}=362.2 L$.
To calculate numerically the partial Fourier transform $ \tilde{\rho}_\ell(k;x,\Omega)$, defined in Eq.~\eqref{eq:partial_FT}, we have used an analogous approach.
For a given pair of $x$ and $\Omega$,  $ \tilde{\rho}_\ell(k;x,\Omega)$ has a form of the type below
\begin{equation}\label{app:partial_FT}
   \tilde{g}(k)=
 \int_{-\infty}^{\infty} d y e^{-i k y} g(y)~.
\end{equation}
We consider the function $g(y)$ in the interval $[-y_{\rm max},y_{\rm max}]$ and select a finite sequence $g_n=g(y_n-y_{\rm max})$, where $y_n=2y_{\rm max} (n/N)$ and $n=0,\ldots,N-1$.
Keeping fixed $y_{\rm max}$, in the limit $N\to \infty$ the discrete values obtained by the DFT of Eq.~\eqref{app:DFT} tend to be
\begin{equation}\label{app:DFT_cont}
\tilde{g}_j = \frac{N}{2y_{\rm max}} \int^{y_{\rm max}}_{-y_{\rm max}} dy e^{-i k_j y} g(y),
\end{equation}
with $k_j= \pi j/ y_{\rm max}$, comparing it with Eq.~\eqref{app:partial_FT}, we find $\tilde{g}(k_j)= \lim_{ y_{\rm max} \to \infty} \lim_{N \to \infty}
\frac{2 y_{\rm max} }{N}\tilde{g}_j$.
%
To calculate $\tilde{\rho}_\ell(k;x,\Omega)$, we have used an approximation of the limit above, taking the finite values of $N=23062$ and $y_{\rm max}=362.2 L$. This enabled us to compute the partial Fourier transform in the range of $[-100/L,100/L]$.
\vspace{2em}

\section{Impossibility of a bound state generated by a non-magnetic Lifshitz impurity}
\label{app:noBoundState}

\begin{figure}[t]
\centering
\begin{overpic}[width=\columnwidth]{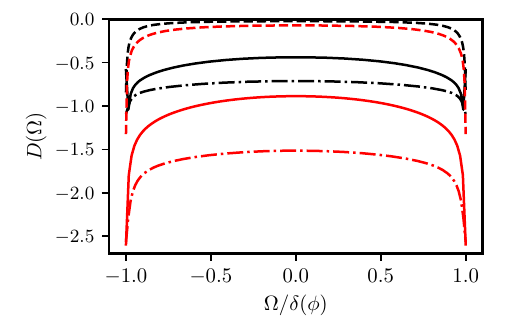}\put(5,55){}\end{overpic}\vspace{0em}
\caption{\justifying (Color online) Discriminant $D(\Omega)$ defined in Eq.~\eqref{app:Discr}, in units of ${\rho^*}^2$ where $\rho^*=1/(\hbar \vD W)$, as a function of the energy $\Omega$ within the mini-gap $\delta(\phi)=\min_k[\epsilon(k,\phi)]$.
Black lines refer to $\mu_0=5 \hbar \vD/L$ and red line refer to  $\mu_0=8 \hbar \vD/L$.
There are three values of the phase difference: $\phi=3\pi/4$ (solid lines), $\phi=0.1\pi$ (dashed lines), and $\phi=0.9\pi$ (dashed-dotted lines).
In all cases, we have a finite short junction length $L/\xi=\pi/20$ and the impurity is placed in the middle of the normal phase stripe at $\bm{r}_d=(0,0)$.
\label{fig:appD}}
\end{figure}

In this Appendix, we show that a single non-magnetic Lifshitz impurity cannot create a bound state in the subgap energy range $|E| <\Delta$.
For a given phase difference $\phi$, within the subgap energy window, the only region in which a generic impurity can create a bound state (with infinite lifetime) is within the minigap, $\delta(\phi)=\min_k[\epsilon(k,\phi)]$.
In particular, the occurrence of a pole of the function $F_{\zeta,j,k,\zeta',j',k'} (\Omega)$, see Eq.~\eqref{app:F}, indicates the emergence of a bound state due to the hybridization of ABSs and the single impurity~\cite{Fetter_book}. This is possible if there is an energy $\Omega$ that solves the following condition
\begin{equation}\label{app:boundC}
{\rm Det}[\openone_\tau-Y(\Omega)]=0~,
\end{equation}
where form of $Y(\Omega)$ depends on the specific nature of the impurity (see App.~\ref{app:ABS-G}).
Here, we focus on the case of a single non-magnetic Lifshitz. Using Eqs.~\eqref{eq:Y-G} and \eqref{eq:X-Lifshitz}, after simple algebraic manipulations, below we report an equivalent equation to Eq.~\eqref{app:boundC} in terms of the $4\times4$ matrix $\bar{G}_0(\boldsymbol{r},\boldsymbol{r}',\Omega)$, defined in Eq.~\eqref{eq:Gbar}, {\rm i.e.}
\begin{widetext}
\begin{equation}{\rm Det}
  \begin{bmatrix}
\frac{1}{u_d A_{\rm c}}- \bar{G}_0(\boldsymbol{r}_d,\boldsymbol{r}_d,\Omega)_{11}-\bar{G}_0(\boldsymbol{r}_d,\boldsymbol{r}_d,\Omega)_{22}
&\bar{G}_0(\boldsymbol{r}_d,\boldsymbol{r}_d,\Omega)_{13}+\bar{G}_0(\boldsymbol{r}_d,\boldsymbol{r}_d,\Omega_0)_{24} \\
\bar{G}_0(\boldsymbol{r}_d,\boldsymbol{r}_d,\Omega)_{31}+\bar{G}_0(\boldsymbol{r}_d,\boldsymbol{r}_d,\Omega)_{42} &-\frac{1}{u_d A_{\rm c}}- \bar{G}_0(\boldsymbol{r}_d,\boldsymbol{r}_d,\Omega)_{33}-\bar{G}_0(\boldsymbol{r}_d,\boldsymbol{r}_d,\Omega)_{44}
\end{bmatrix}=0.
\label{app:boundC2}
\end{equation}
Conversely, we see that for any energy $\Omega$ within the mini-gap $|\Omega|<\delta(\phi)$, Eq.~\eqref{app:boundC2} is an algebraic quadratic equation where the unknown variable is $1/(u_d A_{\rm c})$. This algebraic equation has at least one real solution only if its associated discriminant $D(\Omega)$ is non negative, $D(\Omega)\ge0$, where
\begin{equation}\label{app:Discr}
\begin{aligned}
D(\Omega)=&[\bar{G}_0(\boldsymbol{r}_d,\boldsymbol{r}_d,\Omega)_{11}+\bar{G}_0(\boldsymbol{r}_d,\boldsymbol{r}_d,\Omega)_{22}+\bar{G}_0(\boldsymbol{r}_d,\boldsymbol{r}_d,\Omega)_{33}+\bar{G}_0(\boldsymbol{r}_d,\boldsymbol{r}_d,\Omega)_{44}]^2\\
&-4[\bar{G}_0(\boldsymbol{r}_d,\boldsymbol{r}_d,\Omega)_{13}+\bar{G}_0(\boldsymbol{r}_d,\boldsymbol{r}_d,\Omega)_{24}][\bar{G}_0(\boldsymbol{r}_d,\boldsymbol{r}_d,\Omega)_{31}+\bar{G}_0(\boldsymbol{r}_d,\boldsymbol{r}_d,\Omega)_{42}]~.
\end{aligned}
\end{equation}
\end{widetext}
Fig.~\ref{fig:appD} illustrates the discriminant $D(\Omega)$ as a function of the energy $\Omega$ within the mini-gap $\delta(\phi)$, for three different values of the phase difference: $\phi=3\pi/4$ (solid lines), $\phi=0.1\pi$ (dashed lines) and $\phi=0.9\pi$ (dashed-dotted lines), and two values of the Fermi level: $\mu_0=5 \hbar \vD/L$ (black lines) and $\mu_0=8 \hbar \vD/L$ (red lines).
In all the cases presented, we have a finite short junction length $L/\xi=\pi/20$, and the impurity is placed in the middle of the normal phase stripe $\bm{r}_d=(0,0)$. Here, the discriminant $D(\Omega)$ is always negative, which confirms that a non-magnetic Lifshitz single impurity cannot generate a bound state within the mini-gap.

\bibliography{literaturea}%

\end{document}